\documentclass[%
 reprint,
superscriptaddress,
 amsmath,amssymb,
 aps,
]{revtex4-1}

\usepackage{graphicx}
\usepackage{dcolumn}
\usepackage{bm}

\usepackage{ulem}

\graphicspath{{Immagini/}}
\PassOptionsToPackage{pdftex}{graphicx}
\usepackage{braket}
\usepackage{color}
\usepackage[caption=false]{subfig}
\usepackage{siunitx}
\usepackage{booktabs}
\usepackage[dvipsnames]{xcolor}
\DeclareMathOperator{\tr}{tr}
\DeclareMathOperator{\sign}{sign}

\begin{document}


\title{Quantum noise in the spin transfer torque effect}

%
%
%
%

\author{Camillo Tassi}
\email{camillo.tassi@gmail.com}
\affiliation{Dipartimento di Matematica e Fisica, Universit\`{a} degli Studi Roma Tre, Via della Vasca Navale 84, 00146 Rome, Italy}

\author{Marco Barbieri}
\email{marco.barbieri.qo@gmail.com}
\affiliation{Dipartimento di Scienze, Universit\`{a} degli Studi Roma Tre, Via della Vasca Navale 84, 00146 Rome, Italy}

\author{Roberto Raimondi}
\email{roberto.raimondi@uniroma3.it}
\affiliation{Dipartimento di Matematica e Fisica, Universit\`{a} degli Studi Roma Tre, Via della Vasca Navale 84, 00146 Rome, Italy}


\begin{abstract}
Describing the microscopic details of the interaction of magnets and spin-polarized currents is key to achieve control of such systems at the microscopic level. Here we discuss a description based on the Keldysh technique, casting the problem in the language of open quantum systems. We reveal the origin of noise in the presence of both field-like and damping like terms in the equation of motion arising from spin conductance.
\end{abstract}

\maketitle

\section{Introduction}

The spin transfer torque (STT) is one of the most studied spintronics effects~\cite{pesin2012spintronics}, in particular due to its applications in storage devices. Indeed, the magnetization direction of a ferromagnetic layer, acting as a bit, can be flipped by means of a spin-polarized current, inducing a torque~\cite{Slonczewski:Current-drivenExcitation,Berger1996}. 

The dynamics of the macroscopic magnetization in presence of a magnetic field is usually described by the  Landau-Lifshitz-Gilbert (LLG) equation, which can be introduced by phenomenological arguments~\cite{landau1935theory,gilbert2004phenomenological}. In the LLG equation, two types of terms are usually considered. The first type includes the torque exerted by the total effective magnetic field (the torque is perpendicular to both the magnetization and the magnetic field), whereas the second type takes care of damping effects (the torque is perpendicular to both the magnetization and to its time derivative). In the context of the STT literature, polarized currents appear as additional terms, which may have both a field-like or a damping-like character, depending whether they act as the torques of the first or second type~\cite{xia2002spin,brataas2001spin,stiles2002anatomy,hankiewicz2007gilbert,ralph2008spin,hankiewicz2008inhomogeneous,tatara2008microscopic,tserkovnyak2009transverse,garate2009nonadiabatic,tatara2018effective}.
In hybrid ferromagnetic-metal systems\cite{Tserkovnyak2005,BRATAAS2006,Hellman2017}, the coupling of the electrical current to the macroscopic degree of freedom of the magnetization is obtained by an exchange interaction. Furthermore, the interface between the ferromagnet and the metal is described by an effective spin-mixing conductance\cite{Braatas2000,brataas2001spin,BRATAAS2006}. Within such approach, the dynamics of the magnetization remains purely quasiclassical and the focus is on the diffusive aspects of the charge and spin dynamics of the free carriers.

In recent years, the advances in fast time-resolved measurements have showed that the magnetization dynamics of a nanomagnet crossed by a polarized current presents a stochastic behaviour at a short time interval~\cite{devolder2008single,tomita2008single,cui2010single,cheng2010nonadiabatic}. In such regime, the arising noise should not be such to disturb the device operation, and it could even be engineered to help the magnet switching, reducing dissipative effects and  heating of the material~\cite{ludwig2017strong}. In this respect, it is fundamental to root the phenomenological quantities in the LLG equation in a microscopic description of the magnet as well as of the current, able to take into account noise and quantum effects in the dynamics.
 
The effort of deriving a microscopic description has been carried out by means of different approaches.  In~\cite{swiebodzinski2010spin}, a model was constructed, based on a tunneling Hamiltonian between two normal metal layers separated by a magnet, adopting the Keldysh formalism~\cite{Keldysh1965} to account for the interaction between the magnet and the spin current. The form of the Hamiltonian contains coupling constants that remain to be determined based on phenomenological considerations. In~\cite{wang2013quantum}, instead, an explicit exchange interaction Hamiltonian 
between free propagating electrons and a localized magnetic impurity is considered. The dynamics is then described by means of the associated scattering matrix, while considering the idealised case of a current made of single electrons arriving at the impurity site separately at given times. This is a powerful model for a single magnet, but it presents some difficulties in extending it to more general instances, including coupling to multiple magnets or the presence of electron-electron interaction.      
 
In this paper we combine the two approaches and adopt the Hamiltonian of Ref.~\cite{wang2013quantum} using the Keldysh technique (see e.g. the book by Rammer~ \cite{rammer2007quantum} for a pedagogical introduction), formulated in the context of functional integrals~\cite{kamenev2009keldysh,kamenev2011field}, to obtain a systematic perturbative expansion around the semiclassical limit represented by the LLG equation. Differently from previous investigations we highlight the presence of both field-like and damping-like noise contributions, 
and connect them directly to the spin-mixing conductance, which is expressed in terms of  the transmission scattering amplitudes for  electrons with opposite spin polarization. This is potentially important for extending our treatment to more complex scattering regions, whose behavior may, nevertheless, be described in terms of the spin-dependent transmission scattering amplitudes.
 
The layout of the paper is the following. In section II we introduce the Hamiltonian of the problem and discuss the application of the Keldysh method to derive an equation of motion for the magnet. In section III a numerical example is studied and compared with existing results obtained in complementary approaches. Finally, conclusions are discussed shortly.

\section{The model Hamiltonian}
\begin{figure}
\centering
{\includegraphics[width= \columnwidth]{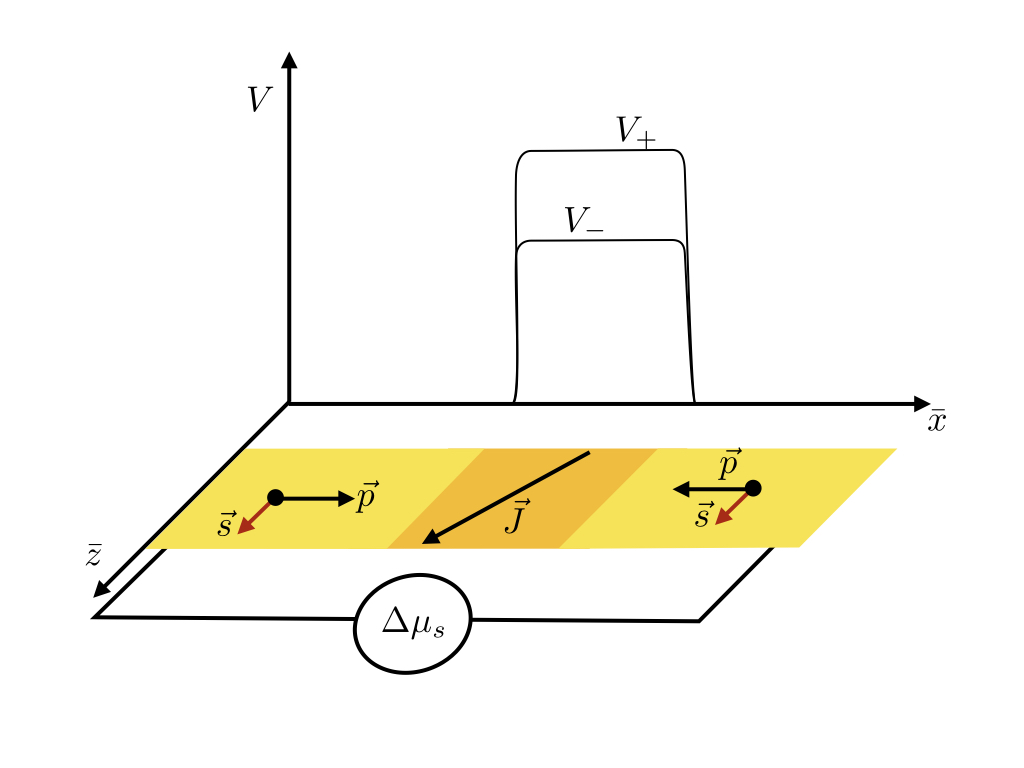}} 
\caption{Conceptual scheme of the device. Electrons move across a magnet with total angular momentum $\vec J$ under the action of a spin potential difference $\Delta \mu_s$. In the magnetic region, they experience a spin-dependent potential $V_+:=d\cdot (\lambda_0+J\, \lambda/2)$ and $V_-:=d\cdot (\lambda_0-J\, \lambda/2)$, depending whether their spin is parallel or anti-parallel to the magnetization vector $\vec J$, where $d$ is the magnet length.}\label{fig:modello}
\end{figure}

The system we model is depicted in Fig.~\ref{fig:modello}. Electrons flow along the $\bar x$ direction across a magnet with total angular momentum $\vec J$ of nanoscopic physical size, yet comprising a large number of constituents  $J\sim10^4$. The electronic current is polarized, due to a spin potential difference $\Delta \mu_s$. The complete Hamiltonian reads~\cite{wang2013quantum}:
\begin{equation}
\label{eqn:WangSham_Hamiltonia}
 H =-\frac{\partial_{\bar{x}}^2}{2\, m} +\gamma\, 
                \vec{B}\cdot \vec{J} +
            \delta(\bar{x})\,
        (\lambda_0 + \lambda\, \vec{J} \cdot
        \vec{s}),
\end{equation}
where the first term is the electron kinetic energy for a particle propagating along a one-dimensional quantum channel, $\vec{B}$ is a weak external magnetic field, $\vec{s}$ is the electron spin. We have taken the approximation that the magnetic region is much smaller than the length of the channel, hence the interaction takes only place at the position $\bar{x}=0$ of the magnet; the actual magnetization is given by $\vec{M}=\gamma\, \vec{J}$, where $\gamma\simeq \lvert e\rvert/m$ is the gyromagnetic ratio of the electron and we have chosen units such that $\hbar=1$. 

A generic observable will evolve in time with the propagator $\hat{U}_{t,t_0}$ from the initial time $t_0$ to an arbitrary time $t$:
 \begin{equation}
   \langle \hat{O}(t)\rangle =
 \tr\left[\hat{U}_{t,t_0}^\dagger\, \hat{O}\, \hat{U}_{t,t_0}\, \hat{\rho}(t_0)\right],
 \label{value}
 \end{equation}
where $\hat{\rho}(t_0)$ is the density matrix for the initial state of the system, and $\hat{U}_{t,t_0}^\dagger=\hat{U}_{t_0,t}$. The Keldysh method thus exploits this symmetry to introduce forward and backward paths in time; the expectation value \eqref{value} is obtained by means of an action $S$ and a function $O$, integrated over bosonic and fermionic degrees of freedom:
\begin{equation}
   \langle \hat{O}(t)\rangle = \int D[\bar{b}_+, \bar{b}_-,\bar{\psi}_+,\bar{\psi}_-,b_+,b_-,\psi_+,\psi_-]
\,  O\, e^{i\, S}
 \end{equation} 
where the complex variable $b_\pm$ and its conjugate $\bar{b}_\pm$ refer to the bosonic field, and the Grassman numbers $\psi_\pm,\, \bar{\psi}_\pm$ refer to the fermionic field. The integration is carried out over the paths $t\to b_\pm(t)$ and  $t\to \psi_\pm(t)$ in both the forward (+) and backward (-) direction of the propagation in time. This formalism produces four propagators in time, two for each time branch and two connecting across them, only three of which are independent. For bosons, the number of propagators is reduced by the Keldysh rotation~\cite{Keldysh1965}:
\begin{subequations}
\label{def:bosons_keldysh_rotation}
\begin{gather}
 b^\text{cl}=\frac{b^+ + b^-}{\sqrt{2}},\qquad b^\text{q}=\frac{b^+ - b^-}{\sqrt{2}},\\
 \bar{b}^\text{cl}=\frac{\bar{b}^+ + \bar{b}^-}{\sqrt{2}},\qquad \bar{b}^\text{q}=\frac{\bar{b}^+ - \bar{b}^-}{\sqrt{2}}.
 \end{gather}
 \end{subequations}
going under the name of  \emph{classical} and \emph{quantum} parts, respectively.

In our case, the fermionic degrees of freedom are the variables of the electrons, and we can treat the magnet spin $J$ as a bosonic field by means of a Holstein-Primakoff (HP) transformation~\cite{holstein1940field}:
\begin{subequations}
\label{def:HolsteinPrimakoff}
\begin{eqnarray}
 &&\hat{J}_+:= J \left( \sqrt{2 - \frac{\hat{b}^\dagger\, \hat{b}}{J}}\right)  \frac{\hat{b}}{\sqrt{J}},\\
 &&\hat{J}_-:= J\, \frac{\hat{b}^\dagger}{\sqrt{J}} \left(  \sqrt{2-\frac{\hat{b}^\dagger\, \hat{b}}{ J}}\right) ,\\
 &&  \hat{J}_z:= J-\hat{b}^\dagger\, \hat{b}
\end{eqnarray}
\end{subequations}
where $\hat{b}^\dagger$ and $\hat{b}$ are bosonic creation and annihilation operators obeying canonical commutation rules. 

\begin{figure}
\centering
{\includegraphics[width= \columnwidth]{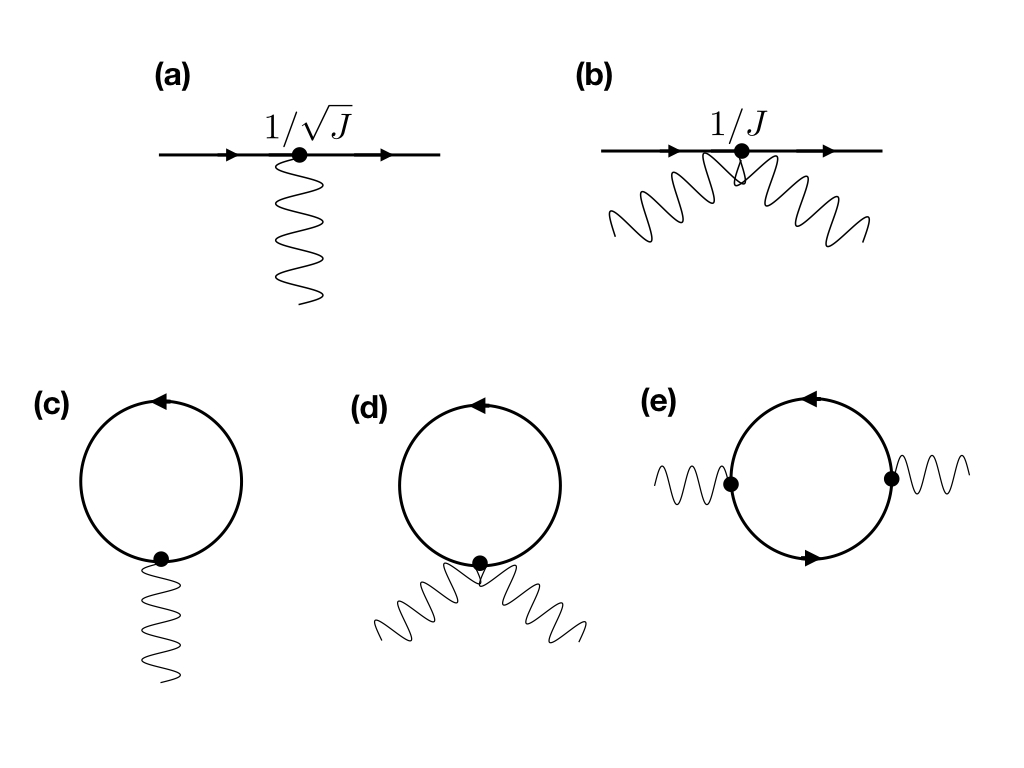}} 
\caption{Feynman vertices and diagrams included in our expansion. (a) single-boson vertex, (b) two-boson vertex, (c) ``tadpole" diagram (d), (e)  boson self-energy diagrams.}
\label{fig:vertici_intro}
\end{figure}

The zero-boson state hence represents the classical limit in which the nano-magnet is perfectly aligned to a given axis, identified with $\hat z$. The presence of boson excitations introduces quantum fluctuations of the magnet spin around this classical axis. In this representation, the interaction between the magnet and the electrons can be depicted by the Feynman vertices in Fig.~\ref{fig:vertici_intro}a and b: the one-boson vertex is of the order $1/\sqrt{J}$, while the two-boson vertex is of the order $1/J$. For a typical nanomagnet~\cite{wang2013quantum}, we can consider only the terms up to the $1/J$-order, that is the Feynman diagrams shown in Fig.~\ref{fig:vertici_intro}c-e. This allows to consider the semiclassical limit of the HP transformation as: 
\begin{subequations}
\label{eqn:J_exp}
\begin{eqnarray}
  &&\hat{J}_{+} = 
  \sqrt{2\, J}\, \hat{b} + O\left( \frac{1}{\sqrt{J}}\right),\\
  && \hat{J}_{-} = 
   \sqrt{2\, J}\, \hat{b}^\dagger +  O\left( \frac{1}{\sqrt{J}}\right),\\
&& \hat{J}_{z} =  J-\hat{b}^\dagger\, \hat{b}.
\end{eqnarray}
\end{subequations}

In order to derive the equation of motion for the magnetization, we have proceeded as follows~\cite{kamenev2009keldysh,kamenev2011field}. The Hamiltonian \eqref{eqn:WangSham_Hamiltonia} has been rewritten taking into account explicitly the linearised HP transformation \eqref{eqn:J_exp}, and three terms have been recognised: $H_{\text{0e}}{=}-\frac{\partial_{\bar{x}}^2}{2\, m} +(\lambda_0 + \lambda\, J s_z)\, \delta(\bar{x})$, which contains the electronic part at fixed magnetic spin, $H_m=\gamma\frac{J}{2}\left(B_+\hat{b}^\dag+B_-\hat{b}\right)+\gamma B_z\left(J-\hat{b}^\dag\hat{b}\right)$ that describes the coupling of $\vec J$ to the magnetic field, and, the interaction term $H_b$, containing all the rest. The Hamiltonians $H_\text{0e}$ and $H_m$ are used to obtain the initial states, then $H_b$ is used as a perturbative correction. The electronic degrees of freedom are traced out, originating the corrections represented by the Feynman diagrams in figures~\ref{fig:vertici_intro}(c-e). We have thus found a functional integral expression for the magnet observables in the form:
 \begin{eqnarray}
   \langle&& \hat{O}(t)\rangle =
 \int D[I_1,I_2]\, e^{-\int dt\, \frac{I_1^2(t)+I_2^2(t)}{2}}\nonumber\\
 &&\times \int D[\bar{b}^{\text{cl}},\bar{b}^{\text{q}},b^{\text{cl}},b^{\text{q}}]\, O(\bar{b}^\text{cl},b^\text{cl})\nonumber\\
 &&\times e^{i\left\lbrace \int dt\, \bar{b}^\text{q}\left[ i\, \partial_t b^\text{cl}+ f(b^\text{cl},\theta,I_1,I_2) \right]+\text{h.c.}\right\rbrace },
  \label{eqn_final_action_intro}
 \end{eqnarray}
where $\theta$ is the angle between the current polarization axis and the magnetization direction $\vec{J}$; the function $f$ is obtained from the sum of the contributions in the Feynman diagrams in figure~\ref{fig:vertici_intro}(c-e). The explicit calculations are reported in the Appendix. Here, $I_1,\, I_2$ are two auxiliary time functions that allows us to linearize the Keldysh action with respect to $b^{\text{q}}$ and $\bar{b}^{\text{q}}$~\footnote
{The linearization is obtained by means of the Hubbard-Stratonovich transformation:
 \begin{equation*}
  e^{-\frac{a}{2}\, x^2}= \sqrt{\frac{1}{2\, \pi\, a}} \int dI\, e^{-\frac{I^2}{2\, a}-i\, x\, I}
 \end{equation*}}.
The semiclassical limit is taken by performing the integration of Eq.~\eqref{eqn_final_action_intro} with respect to $b^{\text{q}},\bar{b}^{\text{q}}$:
 \begin{widetext}
\begin{equation}
 \label{eqn_final_action_intro_b}
   \langle \hat{O}(t)\rangle =
 \int D[I_1,I_2]\, e^{-\int dt\, \frac{I_1^2(t)+I_2^2(t)}{2}}
 \int D[\bar{b}^{\text{cl}},b^{\text{cl}}]\,
 O(\bar{b}^\text{cl},b^\text{cl})
 \delta\left[  i\, \partial_t b^\text{cl}+ f(b^\text{cl},\theta,I_1,I_2) \right] \delta\left[ - i\, \partial_t \bar{b}^\text{cl}+ \bar{f}(b^\text{cl},\theta,I_1,I_2) \right],
 \end{equation}
 \end{widetext}
where $\delta$ is the Dirac function. This implies that 
\begin{equation}
\label{eqn:motion_O_cl}
\begin{cases}
i\, \partial_t b^\text{cl}+ f(b^\text{cl},\theta,I_1,I_2)=0,\\
-i\, \partial_t \bar{b}^\text{cl}+ \bar{f}(b^\text{cl},\theta,I_1,I_2)=0
\end{cases}
\end{equation}
are the equations of the motion. The generic function $t\mapsto I_i(t)$ in the functional integral is weighted by the factor $e^{\int  -\frac{I_1^2(t)+I_2^2(t)}{2} \, dt}$, as in the Martin-Siggia-Rose action~\cite{Martin1973}: the weight is the multivariate Gaussian distribution probability, with zero mean value and unitary variance:
\begin{equation}
 \langle I_i(t)\rangle=0, \qquad \langle I_i(t_1)\, I_j(t_2)\rangle=\delta(t_1-t_2)\, \delta_{ij},
 \label{eqn:average_variance_Langevin}
\end{equation}
that is, $I_i$ must be considered as Langevin terms in the equation of motion. The presence of these  stochastic terms is not surprising, since it is the typical situation of the open quantum systems: when some degrees of freedom are traced over, a stochastic behaviour appears.

Finally, the equation of motion for the magnet can be cast in the LLG form, where the coefficients are expressed in terms of microscopic quantities:
\begin{equation}
 \partial_t \vec{J} = \gamma\, \vec{B}\times\vec{J}+
 C_F\, \hat{z}' \times \vec{J} + 
C_D\, \vec{J}\times (\hat{z}'\times \vec{J})
 \label{eqn:dynamics_equation_intro}
\end{equation}
where $\hat{z}'$ is the  polarization axis of the incoming electrons: $\hat{z}' \times \vec{J}$ and $\vec{J}\times (\hat{z}'\times \vec{J})$ are a field-like and a damping-like term (compare with LLG equation), that produce respectively a precession around the polarization current direction $\hat{z}'$ and an alignment to it. The corresponding microscopic coefficients $C_F$ and $C_D$ are given by:
\begin{subequations}
\begin{eqnarray}
 && C_F = \Re C_1 + \frac{-\cos\theta\, \Im C_2\, I_1 + \Re C_2\, I_2}{\sin\theta} ,\\
 && C_D = \frac{\Im C_1}{J}  +  \frac{\cos\theta\, \Re C_2\, I_1 + \Im C_2\, I_2}{\sin\theta\, J},
\end{eqnarray}
\end{subequations}
where the $C_i$ depend on the scattering matrix and increase with the spin potential difference $\Delta\mu_s$. In particular, $C_1$ is the contribution of the tadpole diagram Fig.~\ref{fig:vertici_intro}(c), while $C_2$ corresponds to the boson self-energy diagrams which are higher order corrections in $1/\sqrt{J}$ Fig.~\ref{fig:vertici_intro}(d-e). This term disappears in the macroscopic limit $J\to +\infty$, while it gives contribution also at zero temperature (both quantum and thermal noise). Comparing our results with the simpler model in~\cite{swiebodzinski2010spin}, we obtained that the scattering of the electrons from the localized magnet results in both field-like and damping-like stochastic correlated terms, as well as a more complex expression for the noise. We may further notice that field-like and damping-like contributions originate from the real (see $\Re C_1$ and $\Re C_2$) and imaginary (see $\Im C_1$ and $\Im C_2$) parts of the coefficients $C_1$ and $C_2$. The presence of the real and imaginary parts is physically due to the different phase shift, upon scattering from the magnet, experienced by electrons with opposite spin orientation.  The microscopic coefficients $C_1$ and $C_2$, whose explicit expression is given in the next Section (see Eqs~\eqref{exp_C1}-\eqref{exp_C2}), are then the result of the interference between the transmission processes for electrons with opposite spin. This interference gives rise to the so-called spin-mixing conductance  \cite{Braatas2000}, which appears whenever paramagnetic conductors are coupled with ferromagnetic metals.

\section{Numerical example}

\label{sec:numerical_example}

We now turn to solving the equation of motion \eqref{eqn:dynamics_equation_intro} we have obtained in the previous section. It is convenient to tackle this by numerical methods. For a comparison to the results in Ref.~\cite{wang2013quantum}, we adopt in our solution the same parameters and initial conditions. In this example, an external magnetic field $\vec{B}$ is only applied at the beginning of the dynamics, and it is assumed that the timescale for the dynamics is much shorter than any thermalization time: for these reasons, the magnetic field and the temperature are relevant only for determining the  initial state. This is the Gibbs ensemble associated to the unperturbed magnet Hamiltonian $H_m$~\cite{wang2013quantum}, characterized by the probability distribution:
\begin{equation}
 P(\hat{z},t=0) = C\, e^{-\beta\, E} = C\, e^{\beta\, \vec{M}\cdot\vec{B}} =
 C\, e^{\beta\, \gamma\,  J\, \hat{z}\cdot\vec{B}} \label{Gibbs_ensemble}
\end{equation}
where $E$ is the magnetic energy and $C$ is a normalization constant $C^{-1}= {4\, \pi\,  \sinh (\beta\,  B\, \gamma\, J)}/({\beta\,  B\, \gamma\, J})$.

We considered a polarized current along the $\hat{z}'$ axis coming from the left to the right: $\Delta\mu_{\text{spin}}^R=0$ and $\Delta\mu_{\text{spin}}^L=:\Delta\mu_{\text{spin}}$.   Under these conditions, we have
\begin{subequations}
\begin{eqnarray}
 && C_1=t_\downarrow(k_F)\, t_\uparrow^\ast(k_F)\frac{\Delta \mu_{\text{spin}}\, \lambda\, m\, }{4\, \pi\,   k_F},\label{exp_C1}\\
 && C_2=t_\downarrow(k_F)\, t_\uparrow^\ast(k_F)\, \lambda\, \sqrt{\frac{  \Delta \mu_{\text{spin}}\,  m}{\pi\, 16\, \epsilon_F}},\label{exp_C2}
\end{eqnarray}
\end{subequations}

where $t_{\uparrow\downarrow}(k_k)$ is the transmission coefficient for the electrons with spin parallel and anti-parallel with respect to $\vec{J}$, respectively, evaluated at the Fermi wavelength (the calculations are reported in the Appendix).
In~\cite{wang2013quantum} it is assumed that $n_e = 1.5\cdot 10^5$ electrons with fixed spin up come from the left to right in a time $t_n$, which is typically of the order of the nanosecond. Since the density current associated to a plane wave $\Psi(\bar{x})=A\, e^{\pm i\, \lvert k\rvert\, \bar{x}}$ is  $I=\frac{1}{m}\, \Im\left( \Psi^\ast\, \partial_{\bar{x}} \Psi\right) = \pm \frac{1}{m}\, \lvert k\rvert\, \lvert A\rvert^2$, we must have $\Delta \mu_{\text{spin}}={2\, \pi\,  n_e}/{t_n}$. The actual numerical values are reported in Table~\ref{table:values_numerical_example}.

\begin{table}
\begin{center}
\begin{tabular}{lccr}
\toprule
Element & Value  & Dimensions \\
\midrule
$t_\uparrow(k_F)$ & $0.067 -0.251\, i$ & \\
$t_\downarrow(k_F)$ & $0.924 -0.265\, i$ & \\
$\Delta\mu_{\text{spin}}$ & $\SI{9.990e-29}{\joule \second}/t_n$  & energy\\
$C_1$ & $(3.312 +5.509\, i)/t_n$  &  time$^{-1}$\\
$C_2$ & $(0.009\, +0.014\, i)/\sqrt{t_n}$  & time$^{-1/2}$\\
$B$ & $\SI{0.05}{\tesla}$  &  magnetic field\\
$\vec{B}$ direction & $(\theta,\phi)=(2.8,1.0)$  & 
\\
$T$ & $\SI{1}{\kelvin}$  & temperature\\
\bottomrule\\
\end{tabular}
\caption{To reproduce the simulation in~\cite{wang2013quantum}, we consider $\lambda_0=3.36\cdot 10^{-28} \si{J.m}$, $\lambda =5.76\cdot 10^{-32} \si{J.m}$, $J=10^4$ and $k_F=13.6\, \si{nm^{-1}}$; furthermore $\Delta \mu_{\text{spin}}=2\, \pi\,  \, 1.5\cdot 10^5/t_n$.}
\label{table:values_numerical_example}
\end{center}
\end{table}

With the adopted choice of the numerical values of the parameters $C_2^2$ is negligible with respect to $C_1$ to a first approximation. We will see that this is inappropriate around the switching time $t \sim t_n$, and the quantum fluctuations become the main contribution to noise.

\begin{figure}
     \includegraphics[width=.8\linewidth]{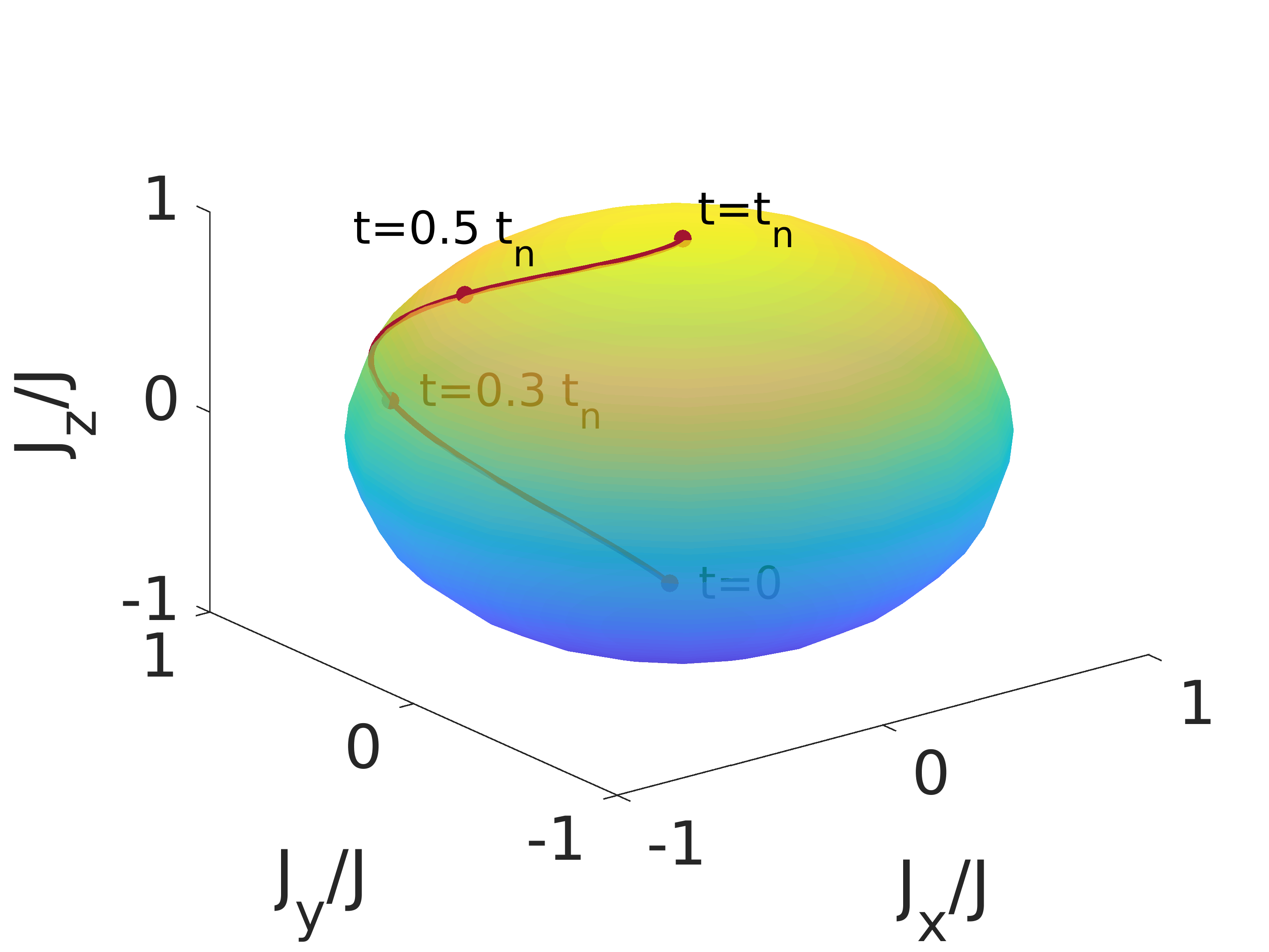}
    \caption[]{Mean value trajectory of $\vec{J}/J$ during the time interval $[0,t_n]$.}
    \label{fig:traiettoria_media_J}
\end{figure}

Fixing $C_2=0$ and choosing the $\theta=0$ axis parallel to the current polarization, the solution of the equation of motion~\eqref{eqn:dynamics_equation_intro} is easily found:
 \begin{equation}
 \begin{cases}
  \theta_S(t,\theta_0)=\theta(t) = 2\, \cot^{-1}\left[\cot \left(\frac{\theta_0}{2}\right)\, e^{\Im C_1\, t}\right],\\
  \phi_S(t,\phi_0)=\phi(t) = \Re C_1\, t + \phi_0,
 \end{cases}
\end{equation}
where $\theta_0$ and $\phi_0$ are the angles for $t=0$. Observe that $\theta(t)$ is a decreasing function that, for $t\to +\infty$, goes to $0$; this represents the damping effect. The trajectory of the average (over  all the possible pairs of initial values $\theta_0$ and $\phi_0$ chosen in the Gibbs ensemble of Eq.~\eqref{Gibbs_ensemble}) value of $\vec J$ is traced in Fig.~\ref{fig:traiettoria_media_J}.

\begin{figure}[bth]
    \subfloat[The initial probability distribution.]
    {\label{fig:iniziale_theta_phi}
     \includegraphics[width=.47\linewidth]{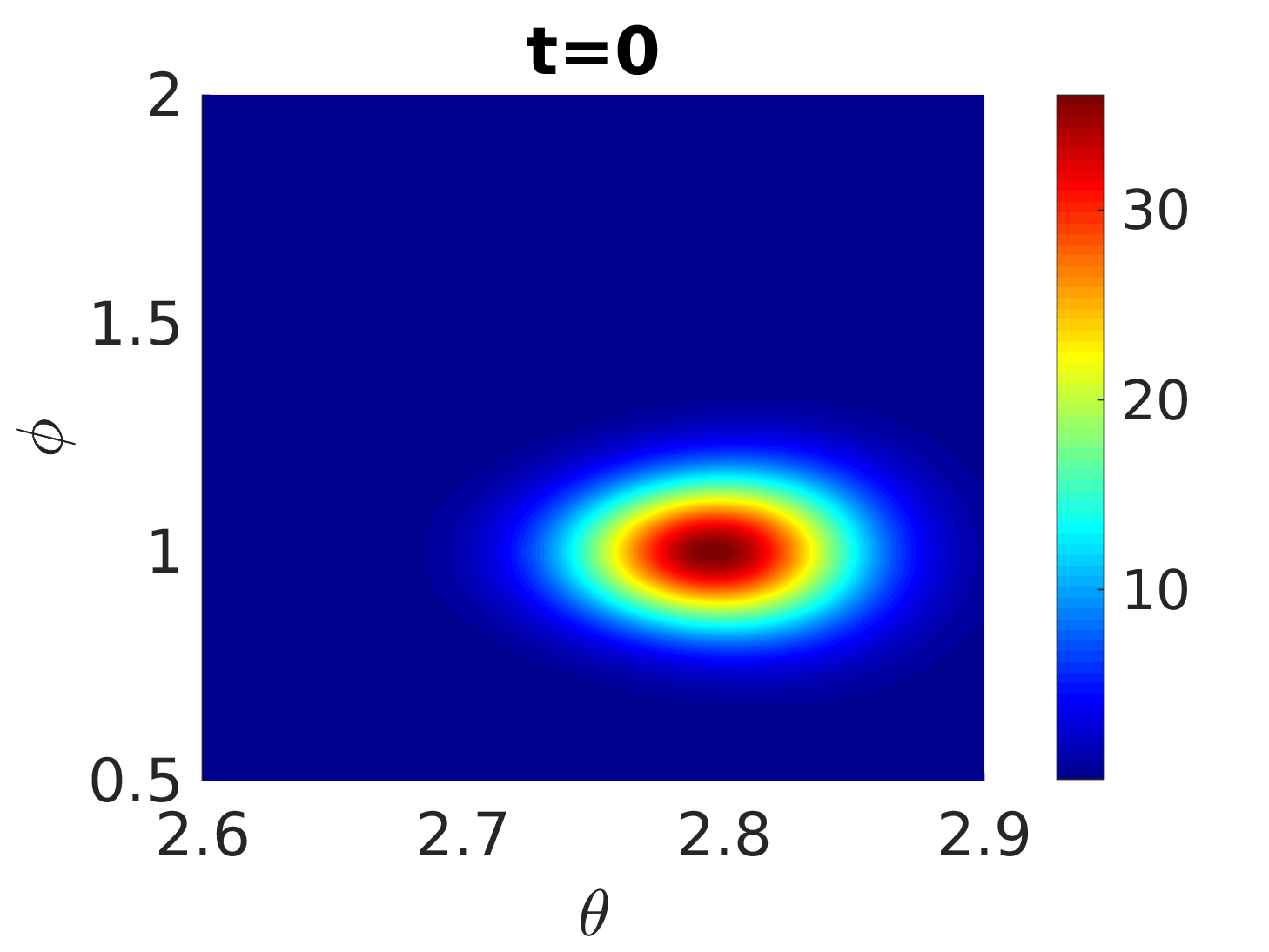}} \quad
    \subfloat[The distribution for $t=0.3\cdot t_n$.]
    {
       \includegraphics[width=.47\linewidth]{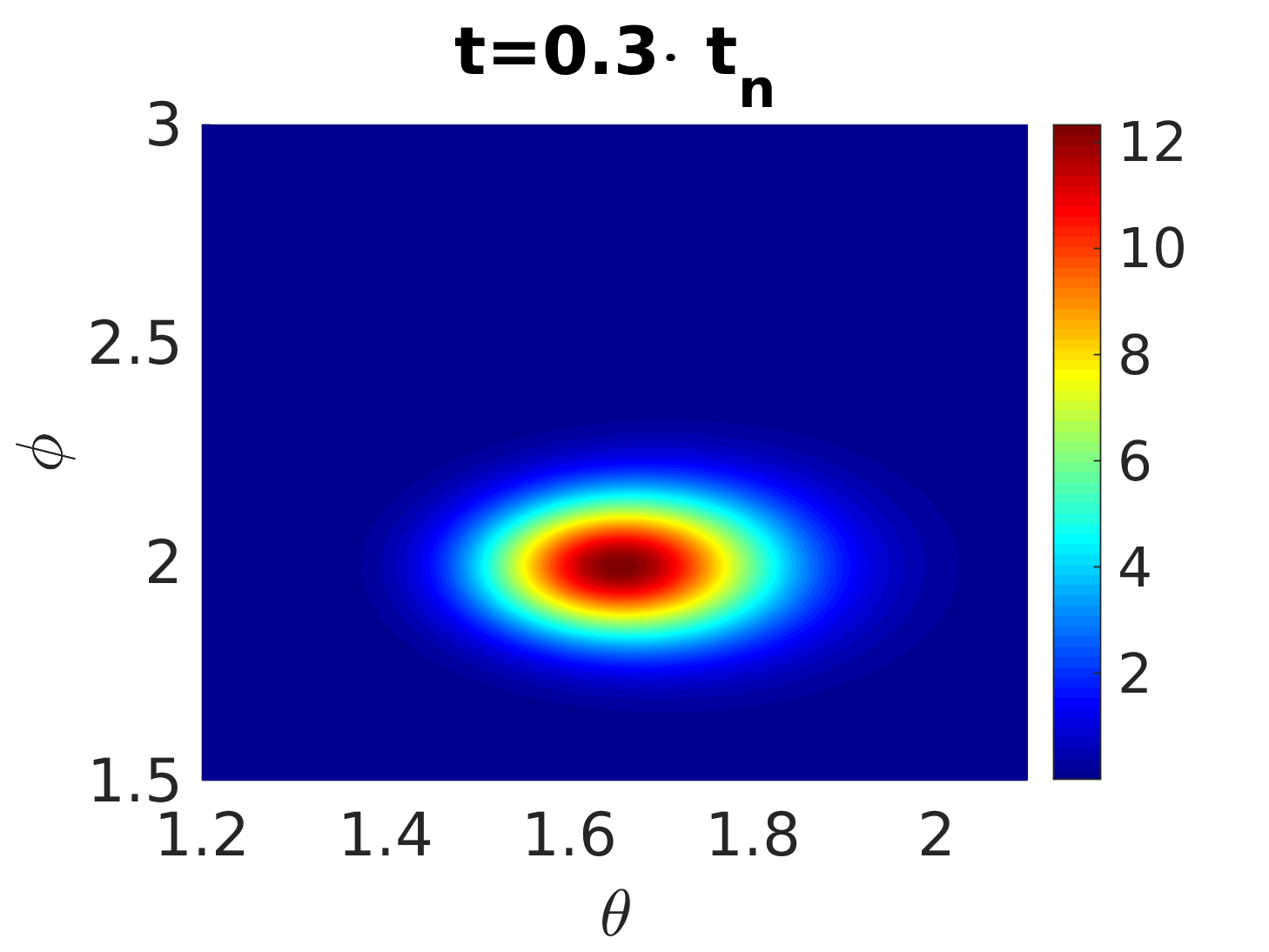}}\\
    \subfloat[The distribution for $t=0.5\cdot t_n$.]
    {
       \includegraphics[width=.47\linewidth]{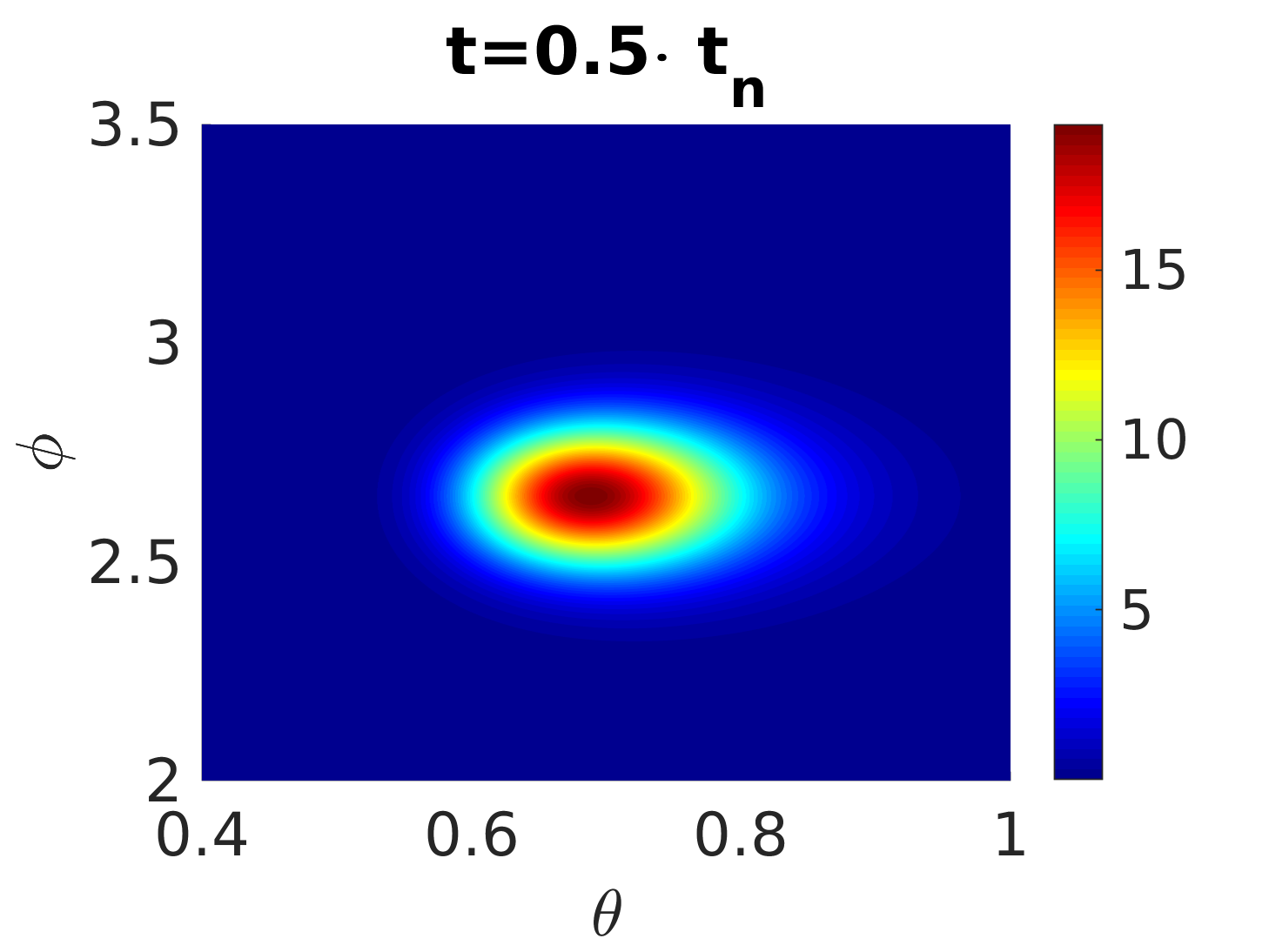}}\quad
    \subfloat[The distribution for $t= t_n$.]
    {\label{fig:t_n_theta_phi}%
       \includegraphics[width=.47\linewidth]{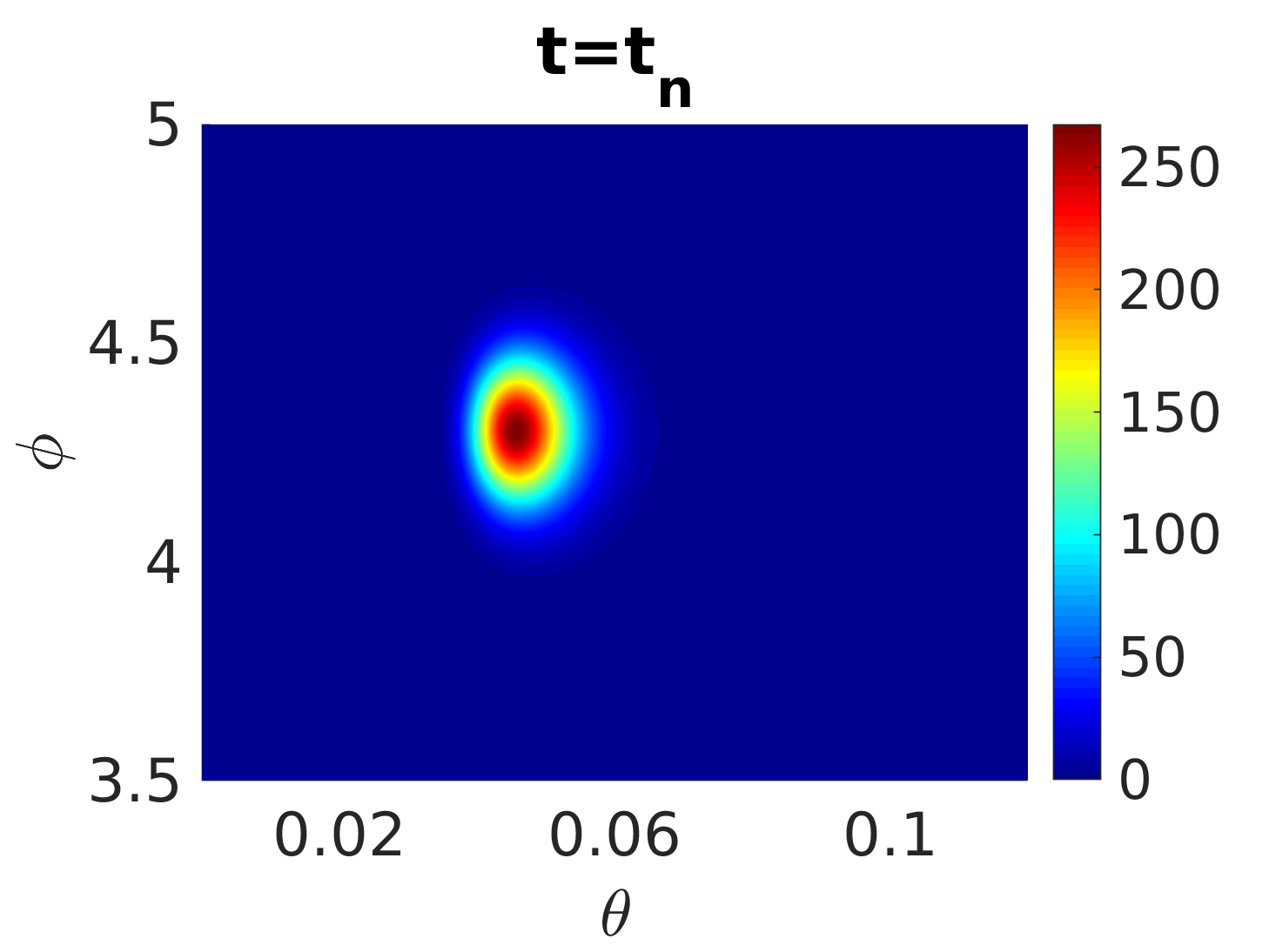}}
    \caption[]{The probability distribution for $(\theta,\phi)$ in radians.}
    \label{fig:distribution_theta_phi}
\end{figure}

We take into account the fluctuations of $J$ in Fig.~\ref{fig:distribution_theta_phi}, where we show the time-dependent probability distribution  $p(\theta,\phi;t)$, with the initial condition $p_0(\theta,\phi)= p(\theta,\phi;t=0)= P(\hat{z},t=0)\, \sin\theta$. The time evolution for $p(\theta,\phi)$ is derived  by considering the evolution of each trajectory:
\begin{eqnarray}
 p(\theta,\phi;t)&=& p_0[\theta_S(-t,\theta),\phi_S(-t,\phi)]\nonumber\\
 &&\times \biggl\lvert \frac{\partial[\theta_S(-t,\theta),\phi_S(-t,\phi)]}{\partial(\theta,\phi)} \biggr\rvert,
\end{eqnarray}
where $\lvert \partial(\theta_S,\phi_S)/\partial(\theta,\phi) \rvert$ is the absolute value of the Jacobian determinant; this expression has been used to obtain the contour plots in Fig.~\ref{fig:distribution_theta_phi}. 

\begin{figure}[t]
    \subfloat[Mean value.]
    {
     \includegraphics[width=.47\linewidth]{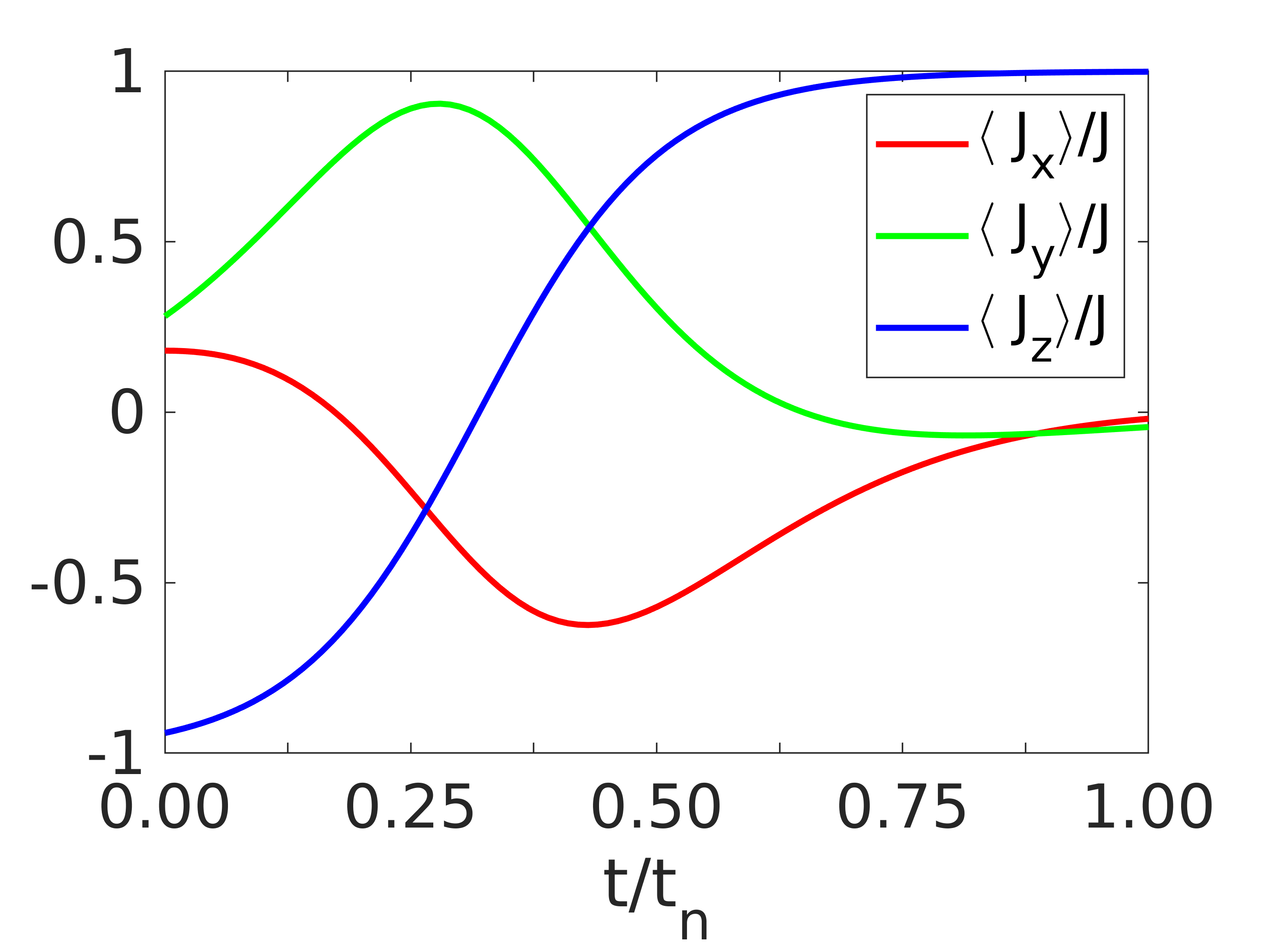}} \quad
    \subfloat[Fluctuation.]
    {\label{fig:fluttuazioni_J}%
       \includegraphics[width=.47\linewidth]{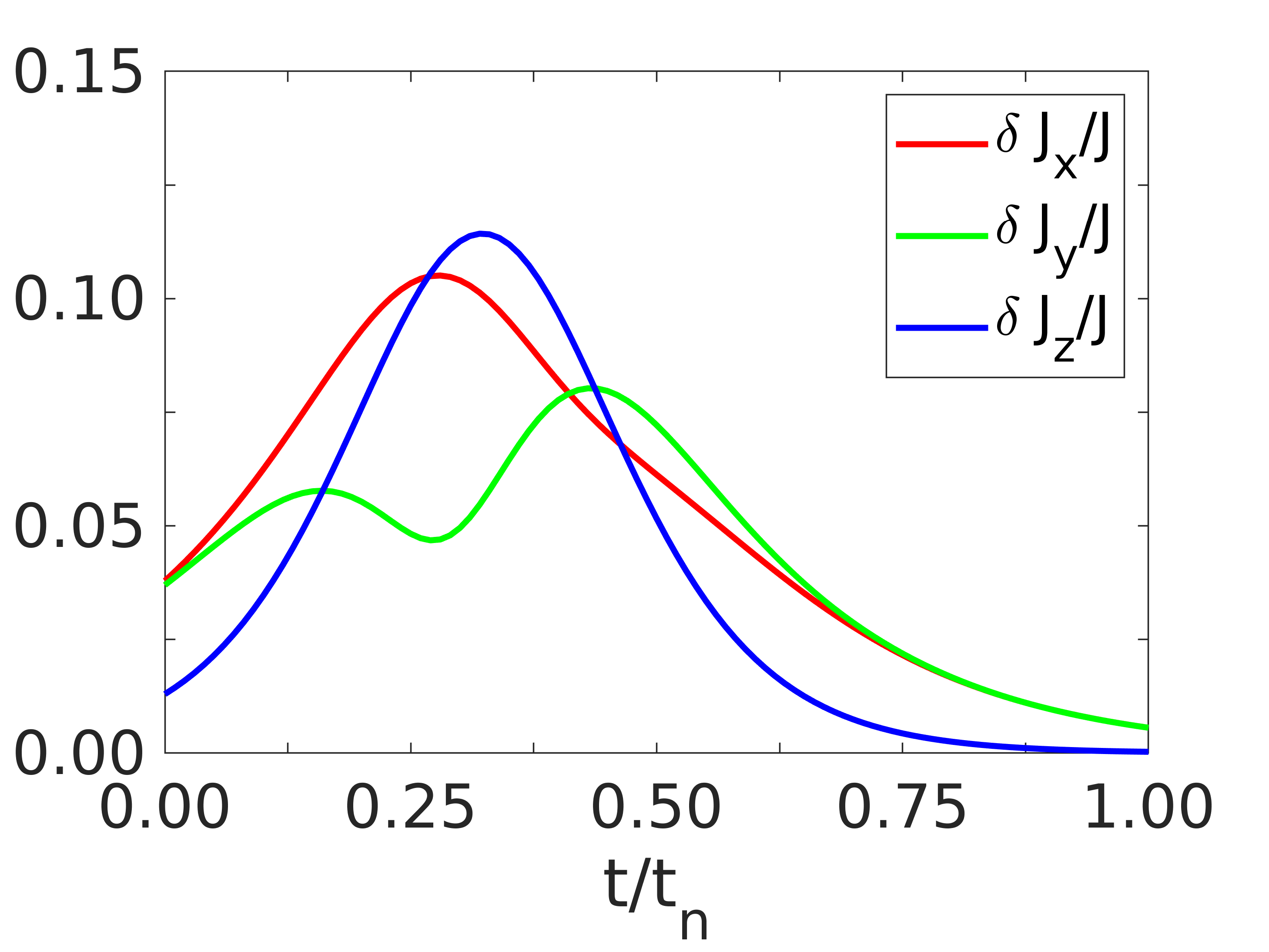}}
    \caption[]{Mean value and standard deviation of $\vec{J}/J$ during the time interval $[0,t_n]$.}
    \label{fig:medie_fluttuazioni_J}
\end{figure}

The mean value and the standard deviation  of the three components of $\vec J$ are summarized in the plots of Fig.~\ref{fig:fluttuazioni_J}, where they are represented as a function of time. As expected based on the values of $C_1$ and  $C_2^2$, these figures are similar to the analogous  ones in Ref.~\cite{wang2013quantum}; in particular, the behaviour of the fluctuations in Fig.~\ref{fig:fluttuazioni_J} is mostly due to the propagation of the initial fluctuations. The only discrepancy in this comparison is the fact that the probability density  in Fig.~\ref{fig:t_n_theta_phi} shows smaller fluctuations. This is not surprising, since for large $t$ damping suppresses all the fluctuations because all the trajectories converge to $\theta=0$: the quantum noise becomes relevant.

As anticipated, quantum fluctuations become important in the long-time limit, in which, due to damping, the angle $\theta(t)$ remains close to $\theta=0$, {\it i.e.} the magnet is almost aligned to the spin of the current. This allows to derive two equations for $\theta$ and $\phi$ from the Eq.~\eqref{eqn:dynamics_equation_intro} as
\begin{subequations}
 \begin{eqnarray}
   &&\theta '(t)=-\Im C_1\, \sin\theta - \left[ I_1(t)\, \Re C_2\, \cos\theta + I_2(t)\, \Im C_2\right]  ,\nonumber\\
   \label{eqn:theta_complete}\\
  &&\phi '(t)= \Re C_1 - \frac{I_1(t)\, \Im C_2\, \cos\theta - I_2(t)\, \Re C_2 }{\sin\theta}.
  \label{eqn:phi_complete}
 \end{eqnarray}
\end{subequations}
The first equation can be linearised with respect to $\theta$ by taking the small angle approximation:
\begin{equation}
\label{eqn:sde_piccolitheta}
 \theta '(t)\simeq
 -\Im C_1\, \theta - I_1(t)\, \Re C_2 - I_2(t)\, \Im C_2.
\end{equation}
Since the last term is a linear combination of two independent Gaussian stochastic processes, it can be cast as a single one with average  $\langle I(t) \rangle=0$ and correlation $ \langle I(t_1)\, I(t_2) \rangle = \lvert C_2\rvert^2\, \delta(t_1-t_2)$. Therefore, Eq.~\eqref{eqn:sde_piccolitheta} reduces to an Ornstein-Uhlenbeck process, whose  corresponding Fokker-Planck equation is:
\begin{equation}
 \partial_t p(\theta,t)=
 \Im C_1\, \partial_\theta[\theta\, p(\theta,t)]+\frac{\lvert C_2\rvert^2}{2}\, \partial_\theta^2 p(\theta,t),
\end{equation}
whose stationary solution is a Gaussian distribution with zero mean value and variance $\lvert C_2\rvert^2/(2\, \Im C_1)$. We refer to the Appendix for the details on the derivation. The variance is of the order $\simeq 5\cdot 10^{-5}$ in our numerical example, thus giving a standard deviation consistent with the difference between our figure~\ref{fig:t_n_theta_phi} and the numerical solution in Ref.~\cite{wang2013quantum}.

The dynamic equation for $\phi$~\eqref{eqn:phi_complete} has a similar structure:
\begin{equation}
 \phi '(t)= \Re C_1 +\tilde{I}(t),
\end{equation}
governed by a stochastic process with $\langle \tilde{I}(t)\rangle=0$, and $\langle \tilde{I}(t_1)\, \tilde{I}(t_2)\rangle=\delta(t_1-t_2)\, \frac{\Im C_2^2\, \cos^2\theta + \Re C_2^2 }{\sin^2\theta}.$ The variance at $\theta=0$ diverges, as one could expect since in this limit the angle $\phi$ is not defined anymore, thus ensuring that the trajectory $t\mapsto (\theta(t),\phi(t))$ of the stochastic process remains continuous. We also remark that in Refs.~\cite{wang2012quantum,wang2013quantum} the diffusion constant for the nano-magnet has been estimated to be of the order of $\SI{e5}{\second^{-1}}$, which is comparable with the thermal noise at $T\sim\SI{3}{\kelvin}$~\cite{wang2012quantum,brown1963thermal}). This is also captured by out treatment: for a flipping time $t_n\sim\SI{1}{\nano\second}$, we find $\lvert C_2\rvert^2\sim\SI{3e5}{\second^{-1}}$.

In our derivation we have assumed that the full system consisting of the electrons and the nano-magnet is not isolated: such interaction with the environment (or with a continuously measuring device) enforces it not to remain in a superposition of different positions~\cite{joos1985emergence,zurek2003decoherence,breuer2002theory}. In principle, however, the Keldysh formalism can be exploited also when relaxing this assumption (refer to the Appendix).

\section{Conclusion}

\label{sec:conclusion}

We have introduced a simple model for the description of noise in STT based on the Keldysh technique. This has allowed us to derive the equation of motion for a nanomagnet interacting with a spin-polarized current; for each term we are able to trace a microscopic origin, and we have made an explicit connection with the spin mixing conductance. We found a good agreement with the model in~\cite{wang2012quantum}, focusing on the scattering matrix approach in the relevant limit.

Thanks to this versatile method, one can extend the treatment to more involved examples, such as those addressing multiple magnets and their correlation that find application in the read/write process, and potential extension to quantum information processing in a solid state architecture.

\begin{acknowledgments}
We thank L.~Mancino and L.~Teresi for discussion.
\end{acknowledgments}

\appendix

\section{The many-body model}

\label{sec:The_many-body_model}

As in Ref.~\cite{wang2013quantum}, we considered electrons only moving in the $\bar{x}$ direction. In principle, the model of Eq.~\eqref{eqn:WangSham_Hamiltonia}  is easily generalizable: if the electric current flows in a device with nanometric transverse dimensions, one can quantize the electron state along $\bar{y}$ and $\bar{z}$ and consider the eigenstates along these directions as current channels~\cite{nazarov2009quantum}. The only complication is that the magnetic scattering center in $\bar{x}=0$ would produces mixing between the channels.

The Keldysh formalism will allow us to treat that model directly in the many-body framework, provided that we  translate the magnet degrees of freedom in terms of boson fields. As in Ref.~\cite{swiebodzinski2010spin}, we will consider the Holstein-Primakoff bosonization defined in Eq.\eqref{def:HolsteinPrimakoff}. By this we consider a semi-classical approximation for the magnet dynamics, in the limit of large $J$ and slight deviation from a coherent state. We then confine to states that are thus combination of few bosons states to ensure that the condition $\langle \hat{J}_{x,y}\rangle\ll J$) holds. In turn, this implies that
\begin{equation}
 \frac{ \langle\hat{b}\rangle}{\sqrt{J}},\, \frac{\langle\hat{b}^\dagger\rangle}{\sqrt{J}} = O\left( \frac{1}{\sqrt{J}}\right),\qquad
 \langle \hat{b}^\dagger\, \hat{b}\rangle \ll J.
\end{equation}
The many-body Hamiltonian $\hat{H} = \hat{H}_{\text{m}} + \hat{H}_{\text{0e}} + \hat{H}_{b}$ is then rewritten in this limit.

The starting point is to identify the scattering states associated to the electronic Hamiltonian $\hat{H}_{\text{0e}}$~\cite{nazarov2009quantum}:
\begin{subequations}
 \label{def:singlescattering}
\begin{eqnarray}
 \Psi&&_{\lvert k\rvert s}(x) = N \nonumber\\
 &&\times
 \begin{cases}
  \Ket{s} e^{i\, \lvert k\rvert\, \bar{x}} + \hat{r}(k) \ket{s} e^{-i\, \lvert k\rvert\, \bar{x}}, & \bar{x}<0,\\
  \hat{t}(k) \Ket{s} e^{i\, \lvert k\rvert\, \bar{x}}, & \bar{x}>0,
 \end{cases} \\
  \Psi&&_{-\lvert k\rvert s}(x) = N\nonumber\\
  &&\times
 \begin{cases}
  \hat{t}(k) \Ket{s} e^{-i\, \lvert k\rvert\, \bar{x}}, & \bar{x}<0,\\
  \Ket{s} e^{-i\, \lvert k\rvert\, \bar{x}} + \hat{r}(k) \Ket{s} e^{i\, \lvert k\rvert\, \bar{x}}, & \bar{x}>0.
 \end{cases}
\end{eqnarray}
\end{subequations}
Here $N$ is a real normalization constant~\footnote
{For example, if we consider that electron are bounded in a region of $\bar{x}$ with dimension $L$ and with periodic boundary conditions, we have $N=1/\sqrt{L}$; if $L$ is much greater with respect to the characteristic electron wave length, we may consider the continuous limit for $k$ and $N=1/\sqrt{2\, \pi}$.
}
and $\Ket{s}$ are spinors. This implies that the transmittivity and reflectivity coefficients are spinor operators in the form
\begin{subequations}
\label{def:spinor_operators_z}
\begin{eqnarray}
 &&\Braket{s_1|\hat{r}(k)|s_2} = 
 \begin{pmatrix}
  r_\uparrow(k) & 0\\
  0 & r_\downarrow(k)
 \end{pmatrix},
 \\
  &&\Braket{s_1|\hat{t}(k)|s_2} = 
 \begin{pmatrix}
  t_\uparrow(k) & 0\\
  0 & t_\downarrow(k)
 \end{pmatrix}
\end{eqnarray}
\end{subequations}
with respect to the $\hat{z}$-quantization axis and
\begin{subequations}
 \begin{eqnarray}
  \label{eqn:rtdef}
 && r_s(k)=\frac{1}{-1+ i\, \frac{\lvert k\rvert}{m\, (\lambda_0+ \lambda\, s\, J/2)}},\\
 && t_s(k) = \frac{1}{1+ i\, \frac{m\, (\lambda_0+\lambda\, s\, J/2)}{\lvert k\rvert}};
 \end{eqnarray}
 \end{subequations}
for convenience we will use the notations $s=\uparrow\downarrow$ and $s=\pm$ with  the same meaning. The states $\Psi_{k s}$  constitute a basis for electrons (respectively coming from left to right and from right to left):
 \begin{equation}
  \hat{H}_{\text{0e}} \Psi_{k s} = \epsilon_{k s}\, \Psi_{k s}, \quad \epsilon_{k s}=\epsilon_k=\frac{k^2}{2\, m},\quad
  s=\pm.
 \end{equation}
 Then the many-electron free Hamiltonian is written as
\begin{equation}
 \hat{H}_{\text{0e}} = \sum_{k s} \epsilon_k\, \hat{c}_{k s}^\dagger\, \hat{c}_{k s},
\label{acca0e}
\end{equation}
where $\hat{c}_{k s}$ creates an electron in the state $\Psi_{k s}$. We stress that the choice to consider an expansion based on the scattering eigenfunctions  $\Psi_{k s}$ is a key technical point to be exploited later on in our discussion.

The interaction term $\hat{H}_b$ containing the $\hat{b},\, \hat{b}^\dagger$ operators can be considered as a perturbation. Indeed, while $\hat{H}_{\text{0e}}$ contains the terms $\lambda_0$ and $\lambda\, J$ (that can be even considered of the similar order -- see Fig.~\ref{fig:modello}), it is easy to check that $\hat{H}_b$ is given by the sum of terms that contain a single bosonic operator ($\hat{b}$ or $\hat{b}^\dagger$) which is of the order $\lambda\, \sqrt{J}$ (and then suppressed by a factor $1/\sqrt{J}$ with respect to $\lambda\, J$) and a term proportional to $\hat{b}\, \hat{b}^\dagger$ which is of the order $\lambda$ (and then suppressed by a factor $1/J$ with respect to $\lambda\, J$).

The incoming current is polarized with respect to an axis denoted $\hat{z}'$ to distinguish it from the one of the magnet $\hat{z}$. The spin states in these two reference frames are related by the rotation
\begin{subequations}
\label{eqn:U_rotation}
\begin{eqnarray}
 &&\hat{U}(\theta,\phi) \Ket{s}=\Ket{s'},\\ 
 &&\Braket{s_1'|\hat{U}|s_2'}=\Braket{s_1|\hat{U}|s_2}\nonumber\\
 && \quad=
 \begin{pmatrix}
  e^{-i\, \phi/2}\, \cos\frac{\theta}{2}    &   -e^{-i\, \phi/2}\, \sin\frac{\theta}{2}\\
  e^{i\, \phi/2}\, \sin\frac{\theta}{2}       &   e^{i\, \phi/2}\, \cos\frac{\theta}{2}  
 \end{pmatrix}
\end{eqnarray}
\end{subequations}
where $\hat{U}=e^{-i\,\hat{s}_{z}\, \phi}\, e^{-i\, \hat{s}_{y}\, \theta}$, with polar and azimuthal angle $\theta$ and $\phi$, respectively. The creation operator $\hat{c}_{k s}^\dagger$ is associated to an electron in the state $\Ket{k,s}$, and similarly for $\hat{c}_{ks'}^\dagger$:
\begin{equation}
 \label{eqn:creationchangebasis}
 \hat{c}^\dagger_{ks_1} = \sum_{s_2'} \Braket{s_2'|s_1} \hat{c}_{ks_2'}^\dagger =
 \sum_{s_2'} \Braket{s_2'|\hat{U}^\dagger|s_1'} \hat{c}_{ks_2'}^\dagger.
\end{equation}
Since the eigenvalues of $H_{\text{0e}}$ do not depend on $s$, we may write the electronic Hamiltonian \eqref{acca0e} in the same form in the two reference frames.

Then we assume that the incoming electrons density matrix is that of a thermal state
\begin{equation}
 \hat{\rho}_0^{s',d} =
 \frac{1}{\mathcal{Z}_{s',d}}\, \exp\left[ -\beta  \sum_k \left( \epsilon_k-\mu^{s',d} \right) \hat{c}_{d\lvert k\rvert, s'}^\dagger\cdot \hat{c}_{d\lvert k\rvert, s'} \right] , 
\end{equation}
where the index $d=\pm$ describes the direction of the electronic motion~\footnote{
This can take into account the action of a potential difference between the left and right regions:
\begin{equation}
 \mu^{s',d} = \epsilon_F + e\, V_0^d + s'\, 2\, \mu_B\, B^d_0,
\end{equation}
where $\epsilon_F$ is the Fermi energy, $V_0^d$ is an electric potential, $\mu_B$ is the Bohr magneton, and $B^d_0$ is a local field due to the presence of  hard ferromagnets layers (see Fig.~\ref{fig:modello}).}.

The operator that annihilates (creates) an electron in $x$ with spin $s$ is given by
\begin{equation}
 \hat{\psi}_{s}(x)=\sum_k  \Psi_{k s}(x;s)\,  c_{k s},\quad
 \hat{\psi}_{s}^\dagger(x)=\sum_k  \Psi_{k s}^\ast(x;s) \, c^\dagger_{k s},
\end{equation}
where $\Psi_{k s}(x;s):=\Bra{s}\Psi_{k s}(x)$, therefore 
\begin{equation}
\hat{\psi}_{s}(0)=N \sum_k  t_s(k)\, c_{k s} ,\quad
\hat{\psi}^\dagger_{s}(0)=N \sum_k  t_s^\ast(k)\, c^\dagger_{k s}. 
\end{equation}
The interaction Hamiltonian can then be written as
\begin{eqnarray}
 \hat{H}_b &=& \lambda \sum_{s_1 s_2}\int d\bar{x}  \Bigl[ -\frac{1}{2}\, \hat{b}^\dagger\, \hat{b}\, \sigma^3_{s_2 s_1} + \frac{\sqrt{2\, J}}{2}\, \sigma^+_{s_2 s_1}\, \hat{b}^\dagger\nonumber\\
 &&+ \frac{\sqrt{2\, J}}{2}\, \sigma^-_{s_2 s_1}\, \hat{b}\Bigr] \hat{\psi}^\dagger_{s_2}(\bar{x})\, \hat{\psi}_{s_1}(\bar{x})\,\delta(\bar{x}),
\end{eqnarray}
where $\sigma^1, \sigma^2, \sigma^3$ are the Pauli matrices, and $\sigma^{\pm}=\sigma^1\pm i\sigma^2$.  This contains a degeneration lifting of the two spin levels, as well as Jaynes-Cummings terms. These describe the situation in which the spin flip of an electron creates or annihilates a bosonic excitation, accounting for the conservation of the total angular momentum. Integration over $\bar{x}$ gives the expression
\begin{eqnarray}
 \hat{H}&&_b=\sum_{s_2' s_1' k_2 k_1} \biggl\lbrace \hat{b}^\dagger\, \hat{b}\, \mathcal{M}_{k_2 k_1}^{s_2 s_1}(\parallel)\, \hat{c}_{k_2 s_2'}^\dagger\,  \hat{c}_{k_1 s_1'} \nonumber\\
 &&+\left[ \hat{b}\, \mathcal{M}_{k_2 k_1}^{s_2 s_1}(\bot)\, \hat{c}_{k_2 s_2'}^\dagger\,  \hat{c}_{k_1 s_1'} + \text{h.c.} \right]  \biggr\rbrace\nonumber\\
 &&=H_{b \parallel}+H_{b \perp},
\end{eqnarray}
where we have also used the basis transformation \eqref{eqn:creationchangebasis}.
The Hamiltonian can be separated into two contributions, depending on the relative orientation with respect to the $\hat{z}$ axis. The coefficient of the parallel contribution is given by
\begin{eqnarray}
 \mathcal{M}^{s_2'\, s_1'}_{k_2\, k_1}(\parallel)&:=&
 -\frac{\lambda}{4}\, N^2 
 \Bigl[  t^\ast_+(k_2)\, t_+(k_1)\, \mathcal{L}_+^{s_2' s_1'}(\parallel)\nonumber\\
 &&+ t^\ast_-(k_2)\, t_-(k_1)\, \mathcal{L}_-^{s_2' s_1'}(\parallel)\Bigr] ,
 \end{eqnarray}
 with
 \begin{subequations}
 \begin{eqnarray}
  &&\mathcal{L}_+^{s_2' s_1'}(\parallel)=
  \begin{pmatrix}
   1+\cos\theta & -\sin\theta\\
   -\sin\theta & 1-\cos\theta
  \end{pmatrix},\\
  &&\mathcal{L}_-^{s_2' s_1'}(\parallel)=
  \begin{pmatrix}
   \cos\theta-1 & -\sin\theta\\
   -\sin\theta & -1-\cos\theta
  \end{pmatrix}.
 \end{eqnarray}
 \end{subequations}
This term is associated to the vertex in Fig.~\ref{fig:vertici_intro}(b). The coefficient for the perpendicular contribution is
 \begin{equation}
 \mathcal{M}^{s_2'\, s_1'}_{k_2\, k_1}(\bot)= \lambda\,
 \frac{\sqrt{2\, J}}{2}\, N^2\, t_-^\ast(k_2)\, t_+(k_1)\, \mathcal{L}^{s_2' s_1'}(\bot),
\end{equation}
with
\begin{equation}
 \mathcal{L}^{s_2' s_1'}(\bot) = e^{-i\, \phi}\,
   \begin{pmatrix}
  \frac{\sin\theta}{2} & -\sin^2\frac{\theta}{2}\\
   \cos^2\frac{\theta}{2}& - \frac{\sin\theta}{2}.
  \end{pmatrix}.
\end{equation}
This term is associated to the vertex in Fig.~\ref{fig:vertici_intro}(a).

\section{Keldysh action}

In the Keldysh formalism~\cite{kamenev2009keldysh}, the magnet-electron action for our system is given by
\begin{eqnarray}
 S&&=\int_{-\infty}^{+\infty} dt \biggl( \bar{b}_+ \, i\, \partial_t\, b_+ + \sum_{k s'} (\bar{\psi}_+)_{k s'} \, i\, \partial_t\, (\psi_+)_{k s'} \nonumber\\
 &&- H[\bar{b}_+,\bar{\psi}_+,b_+,\psi_+] \biggr)  \nonumber\\
 &&-\int_{-\infty}^{+\infty} dt \biggl( \bar{b}_- \, i\, \partial_t\, b_- + \sum_{k s'} (\bar{\psi}_-)_{k s'} \, i\, \partial_t\, (\psi_-)_{k s'} \nonumber\\
 &&- H[\bar{b}_-,\bar{\psi}_-,b_-,\psi_-] \biggr)=
 \int dt\, \mathcal{L}
\end{eqnarray}
where $b_\pm$ are numbers that correspond to the bosonic degrees of freedom and $\psi_\pm$ are  Grassmann numbers for fermions modes. The Keldysh rotation can be applied to reduce the number of propagators: we apply the rotation~\eqref{def:bosons_keldysh_rotation} for bosons, while for fermions we use the Larkin-Ovchinnikov notation~\cite{kamenev2009keldysh}:
\begin{subequations}
\begin{eqnarray}
 &&\psi_1=\frac{\psi_+ + \psi_-}{\sqrt{2}},\qquad \psi_2=\frac{\psi_+ - \psi_-}{\sqrt{2}},\\
 &&\bar{\psi}_1=\frac{\bar{\psi}_+ - \bar{\psi}_-}{\sqrt{2}},\qquad \bar{\psi}_2=\frac{\bar{\psi}_+ + \bar{\psi}_-}{\sqrt{2}}.
\end{eqnarray}
\end{subequations}

The action is the sum of four components: $S_m$, $S_{\text{0e}}$, $S_{b \parallel}$, $S_{b\perp}$. The first term of the Lagrangian $\mathcal{L}$ contains terms associated to the magnet only: 
 \begin{equation}
 \mathcal{L}_m=  \bar{b}_+ \, i\, \partial_t\, b_+-H_m[\bar{b}_+,b_+]-\bar{b}_- \, i\, \partial_t\, b_-+H_m[\bar{b}_-,b_-],
 \end{equation}
which yields the action:
\begin{eqnarray}
 S&&_m=\int dt\,
 \begin{pmatrix}
  \bar{b}^{\text{cl}} & \bar{b}^{\text{q}}
 \end{pmatrix}
\begin{pmatrix}
 0 & i\, \partial_t + \gamma\, B_{z}\\
 i\, \partial_t+ \gamma\, B_{z} & 0
\end{pmatrix}
\begin{pmatrix}
 b^{\text{cl}}\\ b^{\text{q}}
\end{pmatrix}\nonumber\\
&&- \int dt \left[ \frac{\gamma\, \sqrt{J}}{\sqrt{2}}\, B_{+} \left( \bar{b}_+ -\bar{b}_- \right) +\frac{\gamma\, \sqrt{J}}{\sqrt{2}}\, B_{-} \left( b_+ - b_- \right)   \right]  \nonumber\\
&&= \int dt \left[ \bar{b}^{\text{q}} \left( i\, \partial_t + \gamma\, B_{z}\right) b^{\text{cl}} -\gamma\, \sqrt{J}\, B_{+}\, \bar{b}^{\text{q}}  \right] + \text{h.c.}, \nonumber\\
 \label{eqn:actionmagnetwithoutcurrent}
\end{eqnarray}
where the first integral has been evaluated by parts.

The purely electronic contribution is written as 
\begin{eqnarray}
\mathcal{L}_{\text{0e}}&=&  \sum_{k s'} (\bar{\psi}_+)_{k s'} \, i\, \partial_t\, (\psi_+)_{k s'}-H_{0e}[\bar{\psi}_+,\psi_+]\nonumber\\
&&-\sum_{k s'} (\bar{\psi}_-)_{k s'} \, i\, \partial_t\, (\psi_-)_{k s'}+H_{0e}[\bar{\psi}_-,\psi_-].\nonumber\\
\end{eqnarray}
For both arms of the Keldysh contours, the Grassman numbers are indexed to take into account the spin, the direction of the electronic motion, and the momentum.  We then introduce:
\begin{eqnarray}
 \bar{\psi}=
  \biggl( &&\begin{matrix}
        \begin{pmatrix}  
        \bar{\psi}_{1 \uparrow'+} & \bar{\psi}_{2 \uparrow'+}
        \end{pmatrix}
        &
        \begin{pmatrix}
        \bar{\psi}_{1 \downarrow'+} & \bar{\psi}_{2 \downarrow'+} 
        \end{pmatrix}
    \end{matrix}\nonumber\\
    &&
    \begin{matrix}
        \begin{pmatrix}  
        \bar{\psi}_{1 \uparrow'-} & \bar{\psi}_{2 \uparrow'-}
        \end{pmatrix}
        &
        \begin{pmatrix}
        \bar{\psi}_{1 \downarrow'-} & \bar{\psi}_{2 \downarrow'-} 
        \end{pmatrix}        
  \end{matrix}\biggr)
  \end{eqnarray}
and
  \begin{equation}
 \psi=  
 \begin{pmatrix}
        \begin{pmatrix}  
        \psi_{1 \uparrow'+} \\ \psi_{2 \uparrow'+}
        \end{pmatrix}
        \\
        \begin{pmatrix}
        \psi_{1 \downarrow'+} \\ \psi_{2 \downarrow'+} 
        \end{pmatrix}\\
        \begin{pmatrix}  
        \psi_{1 \uparrow'-} \\ \psi_{2 \uparrow'-}
        \end{pmatrix}
        \\
        \begin{pmatrix}
        \psi_{1 \downarrow'-} \\ \psi_{2 \downarrow'-} 
        \end{pmatrix}        
  \end{pmatrix},
\end{equation}
where $\uparrow'\downarrow'$ refer to spin along $\hat{z}'$, $\pm$ to the electron motion direction and each $\psi_{i s' d}$ is a block indexed by the momentum $\lvert k \rvert$:
\begin{subequations}
 \begin{eqnarray}
  &&\bar{\psi}_{i s' \pm}=
  \begin{pmatrix}
   \bar{\psi}_{i s' \pm \lvert k_1\rvert} & \bar{\psi}_{i s' \pm \lvert k_2\rvert} & \cdots
  \end{pmatrix},\\
  &&\psi_{i s' \pm}=
  \begin{pmatrix}
   \psi_{i s' \pm \lvert k_1\rvert} \\ \psi_{i s' \pm \lvert k_2\rvert} \\ \vdots
  \end{pmatrix}.
 \end{eqnarray}
 \end{subequations}
 With this notation, the electronic action is written as
 \begin{eqnarray}
 S_{\text{0e}}= \int &&dt\, \sum_{ks} 
  \begin{pmatrix}
  \bar{\psi}_{1 k s} & \bar{\psi}_{2 k s}
  \end{pmatrix}
  \begin{pmatrix}
   i\, \partial_t - \epsilon_k  &  0\\
   0                            &  i\, \partial_t - \epsilon_k
  \end{pmatrix}\nonumber\\
  &&\times
    \begin{pmatrix}
  \psi_{1 k s} \\ \psi_{2 k s}
  \end{pmatrix}
  = \int dt\, \bar{\psi}\, \check{G}^{-1}_0 \psi,
 \end{eqnarray}
 where
 \begin{equation}
  \check{G}^{-1}_0 = \hat{I}_4 \otimes \hat{G}_0^{-1} \otimes \hat{\gamma}^{\text{cl}}
 \end{equation}
 with $\hat{I}_4$ the $4\times 4$-identity matrix and we use the standard notation:
 \begin{subequations}
 \begin{eqnarray}
  &&\hat{\gamma}^\text{cl} =
  \begin{pmatrix}
   1 & 0\\
   0 & 1
  \end{pmatrix},
 \qquad
  \hat{\gamma}^\text{q} =
  \begin{pmatrix}
   0 & 1\\
   1 & 0
  \end{pmatrix},\\
  &&\left( \hat{G}_{0}^{-1}(t)\right)_{\lvert k_1\rvert \lvert k_2\rvert} =  \delta_{12} \left( i\, \partial_t-\epsilon_{k_1}\right),
 \end{eqnarray}
 \end{subequations}
 where for simplicity we wrote $\delta_{12}$ instead of $\delta_{\lvert k_1\rvert \lvert k_2\rvert}$. We remark how the classical gamma matrix is diagonal in the Keldysh space, while the quantum gamma matrix flips the Keldysh components.

 Finally, we consider the interaction terms, distinguishing between the parallel and perpendicular contributions. From $H_{b \parallel}$ we obtain
 \begin{eqnarray}
  S&&_{b\parallel}= 
  -\int dt\, \sum_{s_2' s_1' k_2 k_1} \frac{1}{2}\, \mathcal{M}_{k_2 k_1}^{s_2' s_1'}(\parallel)
  \bigl[ \left( \bar{b}^{\text{cl}}\, b^{\text{cl}} + \bar{b}^{\text{q}}\, b^{\text{q}}\right)\nonumber\\ 
  &&\times\left( \bar{\psi}_{1 k_2 s_2'}\, \psi_{1 k_1 s_1'}+ \bar{\psi}_{2 k_2 s_2'}\, \psi_{2 k_1 s_1'}\right)\nonumber\\
  &&+\left( \bar{b}^{\text{cl}}\, b^{\text{q}} + \bar{b}^{\text{q}}\, b^{\text{cl}}\right) \left( \bar{\psi}_{2 k_2 s_2'}\, \psi_{1 k_1 s_1'}+ \bar{\psi}_{1 k_2 s_2'}\, \psi_{2 k_1 s_1'}\right)
  \bigr] =\nonumber\\
  &&= -\frac{1}{2}\, \int dt\, \bar{\psi} \biggl[ \left( \bar{b}^{\text{cl}}\, b^{\text{cl}} + \bar{b}^{\text{q}}\, b^{\text{q}}\right) \hat{\mathcal{M}}(\parallel)\otimes\hat{\gamma}^{\text{cl}} \nonumber\\
  &&+ \left( \bar{b}^{\text{cl}}\, b^{\text{q}} + \bar{b}^{\text{q}}\, b^{\text{cl}}\right) \hat{\mathcal{M}}(\parallel)\otimes\hat{\gamma}^{\text{q}}  \biggr] \psi,
 \end{eqnarray}
 where
 \begin{equation}
  \hat{\mathcal{M}}(\parallel):=
  \begin{pmatrix}
   \hat{\mathcal{M}}^{\uparrow'\uparrow'}_{++}(\parallel)  &  \hat{\mathcal{M}}^{\uparrow'\downarrow'}_{++}(\parallel)  & \hat{\mathcal{M}}^{\uparrow'\uparrow'}_{+-}(\parallel)  &  \hat{\mathcal{M}}^{\uparrow'\downarrow'}_{+-}(\parallel)\\
   \hat{\mathcal{M}}^{\downarrow'\uparrow'}_{++}(\parallel)  &  \hat{\mathcal{M}}^{\downarrow'\downarrow'}_{++}(\parallel)  & \hat{\mathcal{M}}^{\downarrow'\uparrow'}_{+-}(\parallel)  &  \hat{\mathcal{M}}^{\downarrow'\downarrow'}_{+-}(\parallel)\\
   \hat{\mathcal{M}}^{\uparrow'\uparrow'}_{-+} (\parallel) &  \hat{\mathcal{M}}^{\uparrow'\downarrow'}_{-+}(\parallel)  & \hat{\mathcal{M}}^{\uparrow'\uparrow'}_{--}(\parallel)  &  \hat{\mathcal{M}}^{\uparrow'\downarrow'}_{--}(\parallel)\\
   \hat{\mathcal{M}}^{\downarrow'\uparrow'}_{-+}(\parallel)  &  \hat{\mathcal{M}}^{\downarrow'\downarrow'}_{-+}(\parallel)  & \hat{\mathcal{M}}^{'\downarrow\uparrow'}_{--}(\parallel)  &  \hat{\mathcal{M}}^{\downarrow'\downarrow'}_{--}(\parallel)
  \end{pmatrix}
 \end{equation}
 and 
 \begin{equation}
  \left( \hat{\mathcal{M}}^{s_2's_1'}_{d_2 d_1}(\parallel)\right)_{\lvert k_2\rvert, \lvert k_1\rvert} :=
  \mathcal{M}^{s_2's_1'}_{d_2\lvert k_2\rvert, d_1\lvert k_1\rvert}(\parallel),\quad d_1,d_2=\pm;
 \end{equation}

 From $H_{b \bot}$ we obtain
 \begin{eqnarray}
  S&&_{b \bot}=-\frac{1}{\sqrt{2}}\nonumber\\
  &&\times\int dt\, \bar{\psi} \left\lbrace  \sum_{\alpha=\text{cl},\text{q}} \left[ b^\alpha\, \hat{\mathcal{M}}(\perp)+ \bar{b}^\alpha\, \hat{\mathcal{M}}^\dagger(\perp)\right]\otimes\hat{\gamma}^\alpha\right\rbrace  \psi,\nonumber\\
 \end{eqnarray}
with analogous meaning of the symbols.

The total action is given by the sum of the four terms above, and can be cast in the form
\begin{eqnarray}
 S&&=S_m+S_{\text{0e}}+S_{b \parallel}+S_{b \perp} 
 \nonumber\\
 &&=
 S_m+\int dt \, 
 \bar{\psi}
    \left[ \check{G}_{0}^{-1} + \check{Q}(\bar{b}^{\text{cl}},\bar{b}^{\text{q}},b^{\text{cl}},b^{\text{q}}) \right] 
      \psi \nonumber
    \\
    &&= S_m+\int dt \, 
 \bar{\psi}
    \left[ \check{G}_{0}^{-1} + \check{Q}_{b \perp}+\check{Q}_{b \parallel} \right] 
      \psi.
\end{eqnarray}

The equation of motion for the magnet is obtained by tracing over the fermionic degrees of freedom:
\begin{eqnarray}
 \int&& \frac{\mathcal{D}[\bar{\psi}\, \psi]}{\prod_{s' d}\tr\left[ \rho_0^{s' d}\right] }\nonumber\\
 &&\times\exp\left\lbrace  i \int dt\, \bar{\psi} \left[ \check{G}_{0}^{-1} +  \check{Q}(\bar{b}^{\text{cl}},\bar{b}^{\text{q}},b^{\text{cl}},b^{\text{q}}) \right]  
         \psi \right\rbrace =\nonumber\\
    =&& \frac{1}{\prod_{s' d}\tr\left[ \rho_0^{s' d}\right]}\, \det\left[ i\,  \left( \check{G}_{0}^{-1} + \check{Q}(\bar{b}^{\text{cl}},\bar{b}^{\text{q}},b^{\text{cl}},b^{\text{q}}) \right)  \right] \nonumber\\
   =&& \det\left[ \check{I} + \check{G}_{0}\, \check{Q}(\bar{b}^{\text{cl}},\bar{b}^{\text{q}},b^{\text{cl}},b^{\text{q}})  \right] 
   \nonumber\\
   =&& e^{\tr\ln\left[ \check{I} + \check{G}_{0} \, \check{Q}(\bar{b}^{\text{cl}},\bar{b}^{\text{q}},b^{\text{cl}},b^{\text{q}})  \right] }
   =e^{i\, S_{m-e}},
\end{eqnarray}
where in the first identity the Gaussian integrals have been used and, in the second, the fact that $\det\left[ i\,  \check{G}_{0}^{-1} \right] = \prod_{s' d}\tr\left[ \rho_0^{s' d}\right]$.

 In the semi-classical limit, we may expand the logarithm:
\begin{eqnarray}
\label{eqn:exp_S_m_e}
 S_{m-e} =&& -i\, \tr\ln\left[ \check{I} + \check{G}_{0}\, \check{Q}(\bar{b}^{\text{cl}},\bar{b}^{\text{q}},b^{\text{cl}},b^{\text{q}})  \right] \nonumber\\
 \simeq&& -i\, \tr\left[ \check{G}_{0}\, \check{Q}(\bar{b}^{\text{cl}},\bar{b}^{\text{q}},b^{\text{cl}},b^{\text{q}})\right] \nonumber\\
 &&+\frac{i}{2}\, \tr\left\lbrace \left[ \check{G}_{0}\,  \check{Q}(\bar{b}^{\text{cl}},\bar{b}^{\text{q}},b^{\text{cl}},b^{\text{q}})\right] ^2\right\rbrace \nonumber\\
 \simeq&& -i\, \tr\left[ \check{G}_{0}\, \check{Q}_{b \perp}\right]  
 -i\, \tr\left[ \check{G}_{0}\, \check{Q}_{b \parallel}\right]\nonumber\\
 && +\frac{i}{2}\, \tr \Bigl[ \check{G}_{0}\,  \check{Q}_{b \perp}\, \check{G}_{0}\,  \check{Q}_{b \perp}\Bigr],
\end{eqnarray}
where we took into account only terms up to the second order in $1/\sqrt{J}$. In particular, the first term is the lowest order term: we have a free electron propagator and a vertex with a single boson; the fermions degrees of freedom are traced over and then we can represent it with the Feynman diagram in Fig.~\ref{fig:vertici_intro}(c). The other two terms are corrections (both of the same order): in particular, in the second term we have a single fermionic line and a two-boson vertex (Fig.~\ref{fig:vertici_intro}(d)), while the third term is composed by two fermionic lines and two single-boson vertices (Fig.~\ref{fig:vertici_intro}(e)).

The Green functions matrix is given by:
\begin{subequations}
\begin{eqnarray}
&&\check{G}_{0}=
 \begin{pmatrix}
  \hat{G}_{0 \uparrow' +}
 & 0 & 0 & 0\\
 0 &
  \hat{G}_{0 \downarrow' +}
 & 0 & 0\\
 0 & 0 &
  \hat{G}_{0 \uparrow' -}
 & 0\\
 0 &  0 & 0 &
  \hat{G}_{0 \downarrow' -}
 \end{pmatrix},\\
  &&\hat{G}_{0 s' d}=
   \begin{pmatrix}
  \hat{G}_{0 s' d}^{R}     &    \hat{G}_{0 s' d}^{K}\\
  0                                            &     \hat{G}_{0 s' d}^{A}
 \end{pmatrix},\quad
 s'=\uparrow',\downarrow',\quad d=\pm,\nonumber\\
\end{eqnarray}
\end{subequations}
where  the four retarded Green functions are equal: $ \hat{G}_{0 \uparrow'\downarrow'\pm}^{R}=: \hat{G}_0^{R}$, with
 \begin{eqnarray}
  \Bigl[ &&\hat{G}_0^{R}(t',t)\Bigr] _{\lvert k_2\rvert  \lvert k_1\rvert} = -i\, \delta_{21}\, \theta(t'-t)\, e^{-i\, \epsilon_{k_1}\, (t'-t)} 
  \nonumber\\
  &&= R_{k_1}(t'-t)\, \delta_{21} 
  \xrightarrow{\text{F.T.}} \left( \epsilon-\epsilon_{k_1} +i\, 0^+ \right)^{-1}\, \delta_{21} ;\nonumber\\
 \label{eqn:generalretardedGreenfunctionsform}
 \end{eqnarray}
   similarly for the advanced Green functions: $\hat{G}_{0 \uparrow'\downarrow'\pm}^{A}=: \hat{G}_0^{A}$, with
 \begin{eqnarray}
  \Bigl[ &&\hat{G}_0^{A}(t',t)\Bigr]_{\lvert k_2\rvert \lvert k_1\rvert} = 
  i\, \delta_{21}\, \theta(t-t')\, e^{-i\, \epsilon_{k_1}\, (t'-t)} \nonumber\\
  &&= A_{k_1}(t'-t)\, \delta_{21} 
  \xrightarrow{\text{F.T.}} \left( \epsilon-\epsilon_{k_1} -i\, 0^+ \right)^{-1} \delta_{21} ,\nonumber\\
   \label{eqn:generaladvancedGreenfunctionsform}
 \end{eqnarray}
   while for $s'=\uparrow',\downarrow'$ and $d=\pm$, the Keldysh Green functions are given by
 \begin{eqnarray}
  \label{eqn:generalKeldyshGreenfunctionsform}
  \Bigl[&& \hat{G}_{0 s' d}^{K}(t',t)\Bigr] _{\lvert k_2\rvert \lvert k_1\rvert} =\nonumber\\
  &&=-i\, \delta_{21}\, \left[ 1-2\, n_F^{s' d}(\epsilon_{k_1})\right]  e^{-i\, \epsilon_{k_1}\, (t'-t)} =
   K_{k_1}^{s' d}(t'-t)\, \delta_{21} \nonumber \\
  &&\xrightarrow{\text{F.T.}} -2\, \pi\, i\,  \delta_{21} \left[ 1-2\, n_F^{s' d}(\epsilon_{k_1})\right] \delta(\epsilon-\epsilon_{k_1}).
 \end{eqnarray}
The Green functions $\hat{G}_{0}^{A,R}$ evaluated in the scattering states have the same simple form as the plane wave functions; the dependence on $(\theta,\phi)$ is contained in the interaction matrices  $\mathcal{M}$ . 

It is useful to observe that 
 we must have the causality condition~\cite{kamenev2009keldysh}
\begin{gather}
 \label{eqn:pureclfield}
 S[\bar{b}^{\text{cl}},b^{\text{cl}},\bar{b}^{\text{q}}=0,b^{\text{q}}=0]=0;
\end{gather}
 in particular we have no linear terms in $b^{\text{cl}}$.

\section{Linear terms in $b$}

\label{sec:Linear_terms_in_b}

The non vanishing linear terms in the $\bar{b},b$-expansion are given by (see Eq.~\eqref{eqn:exp_S_m_e}):
\begin{eqnarray}
 S&&_1=
  -i\, \tr\left[ \check{G}_{0}\, \check{Q}_{b \perp}\right]=\nonumber\\
  &&=
 -i\, \tr\left[ \check{G}_{0} \left\lbrace -\frac{1}{\sqrt{2}} \left[ b^{\text{q}}\, \hat{\mathcal{M}}(\perp)+\bar{b}^{\text{q}}\, \hat{\mathcal{M}}^\dagger(\perp) \right]\otimes\hat{\gamma}^{\text{q}}  \right\rbrace \right] =\nonumber\\
 &&=\frac{i}{\sqrt{2}}\, \sum_{s' d}\tr\left[ b^{\text{q}}\, G^K_{0 s' d}\, \mathcal{M}^{s' s'}_{d d}(\perp) \right] + \text{h.c.} =\nonumber\\
 &&= \frac{1}{\sqrt{2}}\,  \int dt\,  b^{\text{q}}\, \sum_{\lvert k_1\rvert \lvert k_2\rvert s' d}  \delta_{21} \left[ 1-2\, n_F^{s' d}(\epsilon_{k_1}) \right]\cdot\nonumber\\
 &&\cdot\lambda\, N^2\, \frac{\sqrt{2\, J}}{2}\, t^\ast_{\downarrow}(\lvert k_2\rvert)\, t_{\uparrow}(\lvert k_1\rvert)\, s\, e^{-i\, \phi}\, \frac{\sin\theta}{2}+ \text{h.c.}  =\nonumber\\
 &&=-\frac{N^2\, \sqrt{J}\, \lambda}{2}\, \int dt\, b^{\text{q}}\,e^{-i\, \phi}\, \sin\theta\cdot\nonumber\\
 &&\cdot
 \sum_{\lvert k\rvert d} \left[ n_F^{\uparrow' d}(\epsilon_k)-n_F^{\downarrow' d}(\epsilon_k) \right]  t^\ast_{\downarrow}(k)\, t_{\uparrow}(k)+ \text{h.c.}=\nonumber\\
  &&=-C_1^\ast \, \sqrt{J}\, \int dt\, b^{\text{q}}\,  e^{-i\, \phi}\, \sin\theta  + \text{h.c.};
  \label{eqn:S1action}
\end{eqnarray}
in the zero temperature and low differential potential limits~\footnote
{In the continuous limit for $k$ (that is the linear dimension of the system along $\bar{x}$ is much greater with respect to the characteristic electron wavelength) and  in the low temperature limit, it is possible to use the Sommerfeld expansion.
In particular
\begin{equation*}
 n_F^\mu(\epsilon)=\theta(\epsilon-\mu)\simeq \theta(\epsilon-\epsilon_F)+ \delta(\epsilon-\epsilon_F)\, (\mu-\epsilon_F),
\end{equation*}
in the zero temperature limit and assuming  that all the chemical potentials have similar values: $\mu^{s' d}\simeq \epsilon_F$
}:
\begin{equation}
 C_1=\frac{N^2\, \lambda\, m\, t^\ast_{\uparrow}(k_F)\, t_{\downarrow}(k_F)}{2\, k_F} \left( \Delta\mu^L_{\text{spin}}+\Delta\mu^R_{\text{spin}} \right).
\end{equation}

In particular the nanomagnet action up to the first order is given by $S_m+S_1$. Then, for this action, by using the relations $J_+/\sqrt{J}= b^{\text{cl}}$, $\frac{J_-}{\sqrt{J}}= \bar{b}^{\text{cl}}$ and $J_z+O\left(\frac{1}{J}\right)=J$,  the equation of motion~\eqref{eqn:motion_O_cl} reads:
\begin{equation}
\label{eqn:limite_classico_grezzo}
 \begin{cases}
  \left( i\, \partial_t +\gamma\, B_z \right) b^{\text{cl}} - \gamma\, \sqrt{J}\, B_+ - C_1\, \sqrt{J}\, e^{i\, \phi}\, \sin\theta = 0,\\
  \text{complex conjugate},
 \end{cases}
\end{equation}
that is:
\begin{eqnarray}
   \partial_t J_{x,y} &=& \gamma\, [\vec{B}\times\vec{J}]_{x,y} + \Re C_1\, [\hat{z}'\times \vec{J}]_{x,y}\nonumber\\ 
   &&+ \frac{\Im C_1}{J}\,  [\vec{J}\times (\hat{z}'\times \vec{J})]_{x,y};
 \label{eqn:classical_first_magnet}
\end{eqnarray}
 this equations are completed by the condition $\vec{J}\cdot\partial_t\vec{J}=0$ (indeed $\hat{J}_z=J$ up to the $1/\sqrt{J}$ order), which gives rise to
\begin{gather}
\label{eqn:motioneq1ord}
 \partial_t \vec{J} = \gamma\, \vec{B}\times\vec{J} + \Re C_1\, \hat{z}'\times \vec{J} + \frac{\Im C_1}{J}\, \vec{J}\times (\hat{z}'\times \vec{J}).
\end{gather}

Observe that, in the limit $\lambda\to 0$, the potential seen by the electrons does not depend on their spins and  $t^\ast_{\uparrow}(k_F)\, t_{\downarrow}(k_F)= \lvert t^\ast_{\uparrow}(k_F)\rvert^2$; in particular the imaginary part of $C_1$ disappears and, for $\vec{B}=0$, the magnet classical motion is simply a precession around the current polarization axis. This is not surprising: in this case the magnet cannot mix the electrons channels (producing, for example, a spin flip on an electron coming from left to right taken from the larger spin population)  and the ``dissipative'' damping-like term disappears.

\section{Quadratic corrections in $b$}

\label{sec:Quadratic_corrections_in_b}

We consider the quadratic corrections in $\bar{b},b$. They are suppressed by a factor $1/\sqrt{J}$ with respect to the linear terms. We can consider corrections up to quadratic terms (see the expansion~\eqref{eqn:J_exp}). 

At this order, we have Feynman diagrams with both one and two fermionic propagators. In particular, the  one propagator term is (see Eq.~\eqref{eqn:exp_S_m_e}):
\begin{equation}
S_{2-1}=
 -i\, \tr\left[ \check{G}_{0}\, \check{Q}_{b \parallel}\right],
\end{equation}
and the two-fermionic propagator term have the form:
\begin{equation}
  S_{2-2}=
 \frac{i}{2}\, \tr \left[ \check{G}_{0}\,  \check{Q}_{b \perp}\, \check{G}_{0}\,  \check{Q}_{b \perp}\right].
\end{equation}

Before describing the calculations of these two terms (see section~\ref{subsec:1_ferm_propag}, \ref{sec:cl-q_two_fermionic_propagator} and~\ref{sec:q-q_two_fermionic_propagator}),  we will show in the next section what kind of corrections they give rise to in the equation of motion~\eqref{eqn:motioneq1ord}.

In particular we will see that they produce, among others, a term that is quadratic in $b^{\text{q}}$. To include it in our dynamics equation we will show that it is mathematically indistinguishable from a linear action provided you include some stochastic terms. This is not surprising from a physical point of view, since,  when we trace over some degrees of freedom, a pure state can be not distinguishable from a mixed state.

\subsection{Corrections to the motion equation}

\label{subsec:Corrections_to_the_motion_equation}

As we will see in the next sections, the term with the single fermionic propagator is of the form
\begin{equation} 
 \label{eqn:S_2-1_contributo_alla_dinamica}
 S_{2-1} = \int dt\, \tilde{B}_z^{2-1}\, \bar{b}^{\text{cl}}\, b^{\text{q}} + \text{h.~c.}.
\end{equation}
Comparing with the magnetic action~\eqref{eqn:actionmagnetwithoutcurrent}, we see that the contribution to the equation of motion of this term can be considered as a correction (which depends on the angle $\theta$ between the magnet and the polarizzazion of the current) to the $z$-component of the external magnetic field. Anyway the form of the Eq.~\eqref{eqn:limite_classico_grezzo} remains unchanged. This equation is valid when we are in a (moving) frame of reference such that the number of bosons is negligeable with respect to $J$. If we assume that the system decoheres in a classical spin coherent state  in a time that is much shorter with respect to the magnet-dynamics typical times, we can consider also $J_z=J$ at any time and then $\vec{J}\cdot\partial_t \vec{J}=0$. This means that the contribution of the term~\eqref{eqn:S_2-1_contributo_alla_dinamica} to the dynamics equation is zero (since it is parallel to $\vec{J}$ at any time).

The two-fermionic propagator action $S_{2-2}$ gives rise to two terms: one with both classical and quantum bosonic legs $S_{\text{cl-q}}$ (evaluated in the section~\ref{sec:cl-q_two_fermionic_propagator}) and one with two quantum legs $S_{\text{q-q}}$ (see section~\ref{sec:q-q_two_fermionic_propagator}).

In particular it turns out:
\begin{eqnarray}
\label{eqn:classico-quanto-azione}
 S_{\text{cl-q}}&&=
 i\, \int dt_1 \int dt_2 
 \begin{pmatrix}
  \bar{b}^{\text{cl}}(t_1) & \bar{b}^{\text{q}}(t_1)
 \end{pmatrix}\nonumber\\
 &&\times
 \begin{pmatrix}
  0 & D^A(t_1-t_2)\\
  D^R(t_1-t_2) & 0
 \end{pmatrix}
  \begin{pmatrix}
  b^{\text{cl}}(t_2) \\ b^{\text{q}}(t_2)
 \end{pmatrix},
\end{eqnarray}
where the $D$-functions depend on the electronic dynamics. From the fermions point of view, they are the spin-spin response functions of the Kubo formula.

In the typical situations the magnet dynamics is much slower than the fermionic dynamics; for example, in the reference~\cite{wang2013quantum} the typical flipping times for the magnet are of the order of the nanosecond, while the typical electrons Fermi energy $\epsilon_F$ is given by some electronvolts, that is the typical frequencies are of the order  of $\epsilon_F\sim \SI{e16}{\second^{-1}}$. As we will see, this means that we can expand $D$ in frequency:
\begin{equation}
 D(\omega) \sim D_0 + \omega\, D_1;
\end{equation}
the first order in $\omega$  gives rise to a Gilbert damping term (the calculation is similar to that proposed in~\cite{swiebodzinski2010spin}) , but it quite suppressed in our assumption and we will not consider it in the following.

The terms $\Im D^{A}_0=\Im D^{R}_0$ give rise to an action of the form:
\begin{equation}
\int dt\, \tilde{B}^{\text{cl-q}}\, \bar{b}^{\text{cl}}\, b^{\text{q}} + \text{h.c.}
\end{equation}
and we can repeat the same considerations for the action~\eqref{eqn:S_2-1_contributo_alla_dinamica}.

The terms $\Re D^{A}_0=-\Re D^{R}_0$ give rise to an action of the form:
\begin{equation}
i\, \int dt\,  \Re D_0^A(\theta) \, \bar{b}^{\text{cl}}\, b^{\text{q}}+ \text{h.c.}
\end{equation}
They produce two terms, $\Re D_0^R\, J_{x}$ and $\Re D_0^R\, J_{y}$, that must be added respectively to the right side of the first and the second equation in~\eqref{eqn:classical_first_magnet}~\footnote
{ Observe that, in our expression~\eqref{eqn:classico-quanto-azione}, $D_0^{A/R}$ is an addend of $-i\,[G^{-1}]^{A/R}$. In particular, the fact that $\Re D^{A}_0=-\Re D^{R}_0$ and $\Im D^{A}_0=\Im D^{R}_0$ guarantees that the action component  $S_{\text{cl-q}}$ is real.}.
In particular, if $\vec{J}\cdot\partial_t \vec{J}=0$, we must have again that this terms are zero. 
Indeed in the moving reference frame we chose, it must be
\begin{equation}
 J_z = J,\qquad
 J_x=J_y=\partial_t J_z =0.
\end{equation}

Finally we consider the $S_{\text{q-q}}$ component of the action; for simplicity we assume the low temperature and differential potential limits, but the generalization is easy. For compactness, we write
\begin{equation}
 c^{\text{q}} =\sqrt{\frac{\pi\, m\, \lambda^2\, J\, N^4}{8} \left( \lvert \Delta\mu_{\text{spin}}^L\rvert + \lvert \Delta\mu_{\text{spin}}^R\rvert \right)}\,  e^{-i\, \phi}\, \frac{t_\uparrow\, t^\ast_\downarrow}{\sqrt{\epsilon_F}}\, b^{\text{q}},
\end{equation}
and, as we will see in the appendix~\ref{sec:q-q_two_fermionic_propagator}, it turns out:
\begin{equation}
 S_{\text{q-q}} = i\, \int dt \left[ 4\, \bar{c^{\text{q}}}\, c^{\text{q}}-\sin^2\theta\, (\bar{c}^{\text{q}}+c^{\text{q}})^2\right].
\end{equation}
This term is not any longer  linear in $b^{\text{q}}$ and then we cannot apply the considerations done in the section~\ref{sec:Linear_terms_in_b} directly. To linearize this term we will use the Hubbard–Stratonovich transformation (you can compare the following calculations with the simpler case in~\cite{swiebodzinski2010spin}).

We have:
\begin{eqnarray}
 e^{i\, S_{\text{q-q}}}&=&e^{- \int dt \left[ 4\, \bar{c^{\text{q}}}\, c^{\text{q}}-\sin^2\theta\, (\bar{c}^{\text{q}}+c^{\text{q}})^2\right] }\nonumber\\
 &=&
 e^{-\frac{1}{2}
 \begin{pmatrix}
  c^{\text{q}} & \bar{c}^{\text{q}}
 \end{pmatrix}
 A
  \begin{pmatrix}
  \bar{c}^{\text{q}} \\ c^{\text{q}}
 \end{pmatrix}
 }
\end{eqnarray}
where
\begin{subequations}
\begin{eqnarray}
 && A= 2
 \begin{pmatrix}
  2 -\sin^2\theta  &  -\sin^2\theta\\
  -\sin^2\theta  &  2 -\sin^2\theta 
 \end{pmatrix}
 \otimes I_t,
 \\
  && A = U^\dagger \left[  4
 \begin{pmatrix}
  \cos^2\theta &   0\\
  0                                 &  1
 \end{pmatrix}
 \otimes I_t  \right]  U,\\
 && U= \frac{1}{\sqrt{2}}
 \begin{pmatrix}
  1 & 1\\
  1 & -1
 \end{pmatrix}
 \otimes I_t
\end{eqnarray}
\end{subequations}
and $I_t$ is the identity over times. Then, if we put 
\begin{subequations}
\begin{eqnarray}
&& c_1:= 2\, \cos\theta \left(  c^{\text{q}}+ \bar{c}^{\text{q}}\right)/\sqrt{2}, \\
&& c_2:= 2 \left(  c^{\text{q}}- \bar{c}^{\text{q}}\right)/(i\, \sqrt{2}), 
\end{eqnarray}
\end{subequations}
we obtain
 \begin{eqnarray}
 e&&^{i\, S_{\text{q-q}}}= 
 e^{-\frac{1}{2} \int dt \left( \lvert c_1\rvert^2 + \lvert c_2 \rvert^2 \right)}
 \nonumber\\
 &&=\int \mathcal{D}[y^\ast,y]\, e^{-\int dt\, \frac{\lvert y_1\rvert^2 + \lvert y_2 \rvert^2 }{2}}\, e^{-\frac{i}{2}\, \int dt \left( \bar{c}_1\, y_1 + \bar{c}_2\, y_2 + \text{h.c.} \right)}\nonumber\\
 &&=
 \int \mathcal{D}[y^\ast,y]\, e^{-\int dt\, \frac{ \lvert y_1\rvert^2 + \lvert y_2 \rvert^2 }{2}}\nonumber\\
 &&\quad\times  e^{-\frac{i}{2}\, \int dt \left[ \frac{4}{\sqrt{2}} \left( \cos\theta\, \Re y_1 + i\, \Re y_2\right) \bar{c}^{\text{q}} + \text{h.c.}\right]} 
 \nonumber\\
 &&=\int \mathcal{D}[I_1,I_2]\, e^{-\int dt \, \frac{I_1^2+I_2^2}{2}}\nonumber\\
 &&\quad\times  e^{- \left[ \left( C_2^\ast\, I_1 + i\, C_2^\ast\, \cos\theta\, I_2\right) \sqrt{J}\, e^{-i\, \phi}\,  b^{\text{q}} + \text{h.c.}\right]} ,
\end{eqnarray}
where in the second equality we used the Hubbard and Stratonovich transformation, in the last equality we integrated over $\Im y_i$ putting $I_i:=\Re y_i$ and 
\begin{equation}
 C_2:=\sqrt{\frac{\pi\, m\, \lambda^2\, N^4}{4\, \epsilon_F} \left( \lvert \Delta\mu_{\text{spin}}^L\rvert + \lvert \Delta\mu_{\text{spin}}^R\rvert \right)}\, t^\ast_\uparrow\, t_\downarrow.
\end{equation}
By comparing with the dynamic equation for $b$~\eqref{eqn:motion_O_cl}, we get immediately the equation of motion for $\vec{J}$~\eqref{eqn:dynamics_equation_intro}.

\subsection{One fermionic propagator}

\label{subsec:1_ferm_propag}

The non-vanishing one propagator term is given by:
\begin{eqnarray}
 S&&_{2-1}=
 -i\, \tr\left\lbrace  \check{G}_{0} \left[ -\frac{1}{2}  \left( \bar{b}^{\text{cl}}\, b^{\text{q}} + \bar{b}^{\text{q}}\, b^{\text{cl}}\right)  \hat{\mathcal{M}}(\parallel) \otimes\hat{\gamma}^{\text{q}}  \right] \right\rbrace\nonumber\\ 
 &&=\frac{i}{2}\, \sum_{s' d}\tr\left[ \left( \bar{b}^{\text{cl}}\, b^{\text{q}} + \bar{b}^{\text{q}}\, b^{\text{cl}}\right) G^K_{0 s' d}\, \mathcal{M}^{s' s'}_{d d}(\parallel) \right] \nonumber\\
 &&=\frac{i}{2} \, \sum_{s' d} \int dt\, \bar{b}^{\text{cl}}\, b^{\text{q}} \tr\left[ G^K_{0 s' d}\, \mathcal{M}^{s' s'}_{d d}(\parallel)  \right] + \text{h.c.}  \nonumber\\ 
 &&=\frac{1}{2}\,  \int dt\, \bar{b}^{\text{cl}}\, b^{\text{q}}\,  \sum_{\lvert k_1\rvert \lvert k_2\rvert s' d}  \delta_{21} \nonumber\\
 &&\quad\times\left[ 1-2\, n_F^{s' d}(\epsilon_{k_1}) \right]\, \mathcal{M}^{s' s'}_{d\lvert k_2\rvert,d\lvert k_1\rvert}(\parallel) + \text{h.c.} =\nonumber\\
 &&=\frac{\lambda\, N^2}{4} \int dt\, \bar{b}^{\text{cl}}\, b^{\text{q}}\,  
 \sum_{\lvert k\rvert d}  \bigl\lbrace  \left[  n_F^{\uparrow' d}(\epsilon_k)-n_F^{\downarrow' d}(\epsilon_k)\right] \nonumber\\
 &&\quad\times\left[ \lvert t_\uparrow(k)\rvert^2 \left( \cos\theta+1 \right) + \lvert t_\downarrow(k)\rvert^2\left( \cos\theta-1 \right)   \right] \nonumber\\
 &&\quad -\left[1-2\, n_F^{\downarrow' d}(\epsilon_k) \right] \left[ \lvert t_\uparrow(k)\rvert^2 -  \lvert t_\downarrow(k)\rvert^2 \right]  \bigr\} +\text{h.c.},\nonumber\\
\end{eqnarray}
where we have used the property
\begin{equation}
 \mathcal{M}^{\uparrow' \uparrow'}_{\lvert k\rvert \lvert k\rvert}(\parallel) = -\mathcal{M}^{\downarrow' \downarrow'}_{\lvert k\rvert \lvert k\rvert}(\parallel)-
 \frac{\lambda}{2}\, N^2 \left[   \lvert t_\uparrow(k)\rvert^2 -  \lvert t_\downarrow(k)\rvert^2 \right] .
\end{equation}

Then, comparing with the magnetic action~\eqref{eqn:actionmagnetwithoutcurrent}, we see that it can be considered a correction to the $z$-component of the external magnetic field. The low temperature and low differential potentials limit can be evaluated easily.

\subsection{cl-q two fermionic propagator}

\label{sec:cl-q_two_fermionic_propagator}

The second order term with two fermionic propagators is:
 \begin{eqnarray}
 S&&_{2-2}=\nonumber\\
 &&=\frac{i}{4}\, \tr\Biggl\lbrace  \check{G}_{0} \left[  \sum_{\alpha_2=\text{q,cl}} \left[ b^{\alpha_2}\, \hat{\mathcal{M}}(\perp)+\bar{b}^{\alpha_2}\, \hat{\mathcal{M}}^\dagger(\perp) \right]\otimes\hat{\gamma}^{\alpha_2}  \right]\nonumber\\
 &&\times \check{G}_{0} \left[  \sum_{\alpha_1=\text{q,cl}} \left[ b^{\alpha_1}\, \hat{\mathcal{M}}(\perp)+\bar{b}^{\alpha_1}\, \hat{\mathcal{M}}^\dagger(\perp) \right]\otimes\hat{\gamma}^{\alpha_1}  \right]
 \Biggr\rbrace.\nonumber\\
\end{eqnarray}

The terms that do not contain at least a $b^{\text{q}}$ or a $\bar{b}^{\text{q}}$ vanish (see e.g.~the causality condition~\eqref{eqn:pureclfield}) and we have:
 \begin{equation}
  S_{2-2}=S_{\text{cl-q}}+S_{\text{q-q}}.
\end{equation}

For the cl-q term, if we write
\begin{equation}
 \check{A}^\alpha :=\left[ b^{\alpha}\, \hat{\mathcal{M}}(\perp)+\bar{b}^{\alpha}\, \hat{\mathcal{M}}^\dagger(\perp) \right]\otimes\hat{\gamma}^{\alpha}=
 \hat{A}^\alpha \otimes \hat{\gamma}^\alpha
\end{equation}
where $\alpha=\text{q,cl}$, we have
\begin{eqnarray}
 S&&_{\text{cl-q}}=
 \frac{i}{4}\, \tr\left[  \check{G}_0\, \check{A}^{\text{cl}} \,   \check{G}_0\, \check{A}^{\text{q}}+   \check{G}_0\, \check{A}^{\text{q}} \,   \check{G}_0\, \check{A}^{\text{cl}}\right] \nonumber\\
 &&=
 \frac{i}{2}\, \tr\left[  \check{G}_0\, \check{A}^{\text{cl}} \,   \check{G}_0\, \check{A}^{\text{q}} \right] \nonumber\\
 &&=
  \frac{i}{2} \sum_{1,2,3,4} \bigl[ \hat{G}_0^R(1,2)\, \hat{A}^{\text{cl}}(2,3)\,  \hat{G}_0^K(3,4)\, \hat{A}^{\text{q}}(4,1) \nonumber\\
  &&\quad+
 \hat{G}_0^K(1,2)\, \hat{A}^{\text{cl}}(2,3)\,  \hat{G}_0^A(3,4)\, \hat{A}^{\text{q}}(4,1) \bigr]   \nonumber\\
 &&=
  i \int dt_1  \int dt_2\, \sum_{a b} b_a^{\text{cl}}(t_1)\, D_{ab}(t_1,t_2)\,  b_b^{\text{q}}(t_2),\nonumber\\
\end{eqnarray}
where e.g.~$1=(s_1',d_1,\lvert k_1\rvert,t_1)$ and $b_a,b_b=b,\bar{b}$. In particular:
\begin{eqnarray}
 D&&_{ab}(t_1,t_2) =\nonumber\\
 &&\frac{1}{2} \sum_{\substack{s_1', d_1, \lvert k_1\rvert, \\ s_2', d_2, \lvert k_2\rvert,\\ 3, 4}}
 \bigl[ \hat{G}^K_0(2,1)\, \hat{\mathcal{M}}_a(\perp;1,3)\, G_0^A(3,4)\, \hat{\mathcal{M}}_b(\perp;4,2) \nonumber\\
 && \quad+  \hat{G}^R_0(2,1)\, \hat{\mathcal{M}}_a(\perp;1,3)\, G_0^K(3,4)\, \hat{\mathcal{M}}_b(\perp;4,2)  \bigr]\nonumber\\
 &&=
 \sum_{\substack{d_1,s_1',\lvert k_1\rvert,\\ s_2',\lvert k_2\rvert}} S(t_1-t_2)\,
 \mathcal{M}_{a \lvert k_1\rvert \lvert k_2\rvert}^{s_1' s_2'}(\perp,t_1)\, \mathcal{M}_{b \lvert k_2\rvert \lvert k_1\rvert}^{s_2' s_1'}(\perp,t_2)  \nonumber\\
 &&\sim
  \sum_{\substack{d_1,s_1',\lvert k_1\rvert,\\ s_2',\lvert k_2\rvert}} S(t_1-t_2)\,
 \mathcal{M}_{a \lvert k_1\rvert \lvert k_2\rvert}^{s_1' s_2'}(\perp)\, \mathcal{M}_{b \lvert k_2\rvert \lvert k_1\rvert}^{s_2' s_1'}(\perp);\nonumber\\
\end{eqnarray}
in the last approximation we used the fact that the fermionic dynamics is much faster than the bosonic one~\footnote
{a similar approximation is made in~\cite{wang2013quantum}, since only one electron scattering per time is considered.} and we have defined:
\begin{eqnarray}
  S&&(t_1-t_2):=\frac{1}{2} \biggl[  K^{s_1'}_{ \lvert k_1\rvert d_1}(t_2-t_1)\,  A_{ \lvert k_2\rvert }(t_1-t_2) \nonumber\\
  &&\quad +
 R_{ \lvert k_1\rvert}(t_2-t_1)\, K^{s_2'}_{ \lvert k_2\rvert d_1 }(t_1-t_2) \biggr] \xrightarrow{\text{F.T.}}\nonumber\\
 &&\frac{1}{2} \int dt\, e^{i\, \omega\, t} \Bigl[  K^{s_1'}_{ \lvert k_1\rvert d_1}(-t)\,  A_{ \lvert k_2\rvert }(t)   +
 R_{ \lvert k_1\rvert}(-t)\, K^{s_2'}_{ \lvert k_2\rvert d_1 }(t)\Bigr]  \nonumber\\
 &&=
 \frac{1}{2} \int \frac{d\epsilon}{2\, \pi} \Bigr[  K^{s_1'}_{ \lvert k_1\rvert d_1}(\epsilon)\,  A_{ \lvert k_2\rvert }(\epsilon+\omega) \nonumber\\
 &&\, +
 R_{ \lvert k_1\rvert}(\epsilon-\omega)\, K^{s_2'}_{ \lvert k_2\rvert d_1 }(\epsilon)\Bigl]
 =
 i\, \frac{n_F^{s_1' d_1}(\epsilon_{\lvert k_1\rvert}) -n_F^{s_2' d_1}(\epsilon_{\lvert k_2\rvert} )}{\epsilon_{\lvert k_1\rvert} - \epsilon_{\lvert k_2\rvert}+\omega-i\, 0^+}\nonumber\\
 &&=
 i \left[ n_F^{s_1' d_1}(\epsilon_{\lvert k_1\rvert}) -n_F^{s_2' d_1}(\epsilon_{\lvert k_2\rvert} )\right] \hat{f}(\epsilon_{\lvert k_1\rvert} - \epsilon_{\lvert k_2\rvert}+\omega)\nonumber\\
\end{eqnarray}
where we used relations~\eqref{eqn:generalretardedGreenfunctionsform}, \eqref{eqn:generaladvancedGreenfunctionsform} and~\eqref{eqn:generalKeldyshGreenfunctionsform}. 
Since we have $[\mathcal{L}(\perp)]^2=0$, the terms that multiplies $b\, b$ and $\bar{b}\, \bar{b}$ disappear and we may adjust the surviving terms to obtain the expression~\eqref{eqn:classico-quanto-azione}. In particular
 the term that multiplies $\bar{b}^\text{cl}\, b^{\text{q}}$ is
 \begin{eqnarray}
  D&&^A(\omega)= \sum_{\substack{d, s_1', \lvert k_1\rvert, \\ s_2', \lvert k_2\rvert}}
  i\, \frac{n_F^{s_1' d}(\epsilon_{\lvert k_1\rvert}) -n_F^{s_2' d}(\epsilon_{\lvert k_2\rvert} )}{\epsilon_{\lvert k_1\rvert} - \epsilon_{\lvert k_2\rvert}+\omega-i\, 0^+}\nonumber\\
  &&\quad\times 
  (\mathcal{M}^\dagger)_{\lvert k_1\rvert \lvert k_2\rvert}^{s_1' s_2'}(\perp)\, \mathcal{M}_{\lvert k_2\rvert \lvert k_1\rvert}^{s_2' s_1'}(\perp) \nonumber\\
  &&=
   i\, \frac{\lambda^2\, J^2\, N^4}{2} \sum_{\substack{d, s_1', \lvert k_1\rvert, \\ s_2', \lvert k_2\rvert}}
   \lvert t_\uparrow(k_1)\, t_\downarrow(k_2)\rvert^2\nonumber\\
   &&\quad\times \frac{n_F^{s_1' d}(\epsilon_{\lvert k_1\rvert}) -n_F^{s_2' d}(\epsilon_{\lvert k_2\rvert} )}{\epsilon_{\lvert k_1\rvert} - \epsilon_{\lvert k_2\rvert}+\omega-i\, 0^+}
   (\mathcal{L}^\dagger)^{s_1' s_2'}(\perp)\, \mathcal{L}^{s_2' s_1'}(\perp).\nonumber\\
    \label{def:DAomega}
 \end{eqnarray}
Instead the term that multiplies $b^\text{cl}\, \bar{b}^{\text{q}}$ is:
 \begin{eqnarray}
 \label{def:DRomega}
  D&&^R(\omega)= \sum_{\substack{d, s_1', \lvert k_1\rvert, \\ s_2', \lvert k_2\rvert}}
  i\, \frac{n_F^{s_1' d}(\epsilon_{\lvert k_1\rvert}) -n_F^{s_2' d}(\epsilon_{\lvert k_2\rvert} )}{\epsilon_{\lvert k_1\rvert} - \epsilon_{\lvert k_2\rvert}-\omega-i\, 0^+}\nonumber\\
  &&\quad\times\mathcal{M}_{\lvert k_1\rvert \lvert k_2\rvert}^{s_1' s_2'}(\perp)\, (\mathcal{M}^\dagger)_{\lvert k_2\rvert \lvert k_1\rvert}^{s_2' s_1'}(\perp) \nonumber\\
  &&=
  i\, \frac{\lambda^2\, J^2\, N^4}{2} \sum_{\substack{d, s_1', \lvert k_1\rvert, \\ s_2', \lvert k_2\rvert}}
   \lvert t_\uparrow(k_1)\, t_\downarrow(k_2)\rvert^2\nonumber\\
  &&\quad\times  \frac{n_F^{s_1' d}(\epsilon_{\lvert k_1\rvert}) -n_F^{s_2' d}(\epsilon_{\lvert k_2\rvert} )}{\epsilon_{\lvert k_1\rvert} - \epsilon_{\lvert k_2\rvert}+\omega+i\, 0^+}\, (\mathcal{L}^\dagger)^{s_1' s_2'}(\perp)\, \mathcal{L}^{s_2' s_1'}(\perp).\nonumber\\
 \end{eqnarray}

We may now assume that $\epsilon\gg\omega$, where $\epsilon$ are the typical electrons energies~\footnote
{
In particular, for sufficiently small values of $\omega$, we have that the the contribution to $S(\omega)$ is non negligeable only for $\epsilon_{\lvert k_1\rvert} \sim \epsilon_{\lvert k_2\rvert}$; but in that case, for  temperatures and differential potentials sufficiently small, $n_F^{s_1' d_1}(\epsilon_{\lvert k_1\rvert}) -n_F^{s_2' d_1}(\epsilon_{\lvert k_2\rvert} )$ is non zero only for $\epsilon_{\lvert k_1\rvert} \sim \epsilon_{\lvert k_2\rvert}\sim\epsilon_F$. So  we have to assume $\epsilon_F\gg \omega$.
}:
\begin{eqnarray}
 S&&(\omega) = i \left[ n_F^{s_1' d_1}(\epsilon_{\lvert k_1\rvert}) -n_F^{s_2' d_1}(\epsilon_{\lvert k_2\rvert} )\right]  \hat{f}(\epsilon_{\lvert k_1\rvert} - \epsilon_{\lvert k_2\rvert}+\omega)\nonumber\\
 &&\sim
 i \left[ n_F^{s_1' d_1}(\epsilon_{\lvert k_1\rvert}) -n_F^{s_2' d_1}(\epsilon_{\lvert k_2\rvert} )\right]\nonumber\\
 &&\quad \times 
 \left[  \hat{f}(\epsilon_{\lvert k_1\rvert} - \epsilon_{\lvert k_2\rvert})+\omega\,  \hat{f}'(\epsilon_{\lvert k_1\rvert} - \epsilon_{\lvert k_2\rvert})\right] \xleftarrow{\text{F.T.}}\nonumber \\
&& i  \left[ n_F^{s_1' d_1}(\epsilon_{\lvert k_1\rvert}) -n_F^{s_2' d_1}(\epsilon_{\lvert k_2\rvert} )\right]
\Bigl[   \delta(t_1-t_2) \,  \hat{f}(\epsilon_{\lvert k_1\rvert} - \epsilon_{\lvert k_2\rvert})\nonumber\\
&&\quad+i\,   \delta'(t_1-t_2) \,  \hat{f}'(\epsilon_{\lvert k_1\rvert} - \epsilon_{\lvert k_2\rvert})\Bigr]=
 S^0 + S^1,
 \label{eqn:q-cl_completa_in_omega}
\end{eqnarray}
and here $S^1$ is the term that contain the first order Dirac delta derivative. 

We concentrate here on the term $S^0$. By using the formula of Sokhotski–Plemelj, it gives rise to:
\begin{eqnarray}
 D&&^0_{ab}(t_1-t_2)=\nonumber\\
 && i \sum_{d_1,s_1', s_2'} \int_0^\infty d\lvert k_1\rvert  \int_0^\infty d\lvert k_2\rvert \left[ n_F^{s_1' d_1}(\epsilon_{\lvert k_1\rvert}) -n_F^{s_2' d_1}(\epsilon_{\lvert k_2\rvert} )\right]\nonumber\\
 &&\quad\times\delta(t_1-t_2) \left[ i\, \pi\, \delta(\epsilon_{\lvert k_1\rvert}-\epsilon_{\lvert k_2\rvert}) + P\frac{1}{\epsilon_{\lvert k_1\rvert}-\epsilon_{\lvert k_2\rvert}} \right] \nonumber\\
 &&\quad\times
 \mathcal{M}_{a \lvert k_1\rvert \lvert k_2\rvert}^{s_1' s_2'}(\perp)\, \mathcal{M}_{b \lvert k_2\rvert \lvert k_1\rvert}^{s_2' s_1'}(\perp) .
\end{eqnarray}
The non vanishing terms are:
\begin{subequations}
 \begin{eqnarray}
  \Re &&D_0^A(t_1-t_2) = -\pi\,  \delta(t_1-t_2) \nonumber\\
  &&\,\times 
  \sum_{d,s_1',s_2'}\int_0^{\infty} d\epsilon \left[ n_F^{s_1' d}(\epsilon) -n_F^{s_2' d}(\epsilon )\right]  \nonumber\\
  &&\,\times 
  \frac{m}{2\,\epsilon} \left( \mathcal{M}^\dagger\right) _{ \lvert k_\epsilon\rvert \lvert k_\epsilon\rvert}^{s_1' s_2'}(\perp)\, \mathcal{M}_{ \lvert k_\epsilon\rvert \lvert k_\epsilon\rvert}^{s_2' s_1'}(\perp) \nonumber\\
  &&=
  -\delta(t_1-t_2)\, \frac{\pi\, \lambda^2\, J\, N^4\, m}{4} \, \cos\theta \nonumber\\
  &&\,\times 
  \sum_d \int_0^{\infty} d\epsilon \left[ n_F^{\uparrow' d}(\epsilon) -n_F^{\downarrow' d}(\epsilon )\right] \frac{\lvert t_\uparrow(k)\,  t_\downarrow(k)\rvert^2}{\epsilon},\nonumber\\ \\
  \Im &&D_0^A(t_1-t_2) = \delta(t_1-t_2)\, \frac{\lambda^2\, J\, N^4\, m}{8} \sum_d \int_0^\infty \frac{d\epsilon_1}{\sqrt{\epsilon_1}}  \nonumber\\
  &&\,\times \int_0^\infty \frac{d\epsilon_2}{\sqrt{\epsilon_2}}
 \biggl\lbrace n_F^{\uparrow' d}(\epsilon_1)+n_F^{\downarrow' d}(\epsilon_1)  \nonumber\\
  &&\,+ \left[ n_F^{\uparrow' d}(\epsilon_1)-n_F^{\downarrow' d}(\epsilon_1) \right] \cos\theta \nonumber\\
  &&\,
 -n_F^{\uparrow' d}(\epsilon_2)-n_F^{\downarrow' d}(\epsilon_2)+ \left[ n_F^{\uparrow' d}(\epsilon_2)-n_F^{\downarrow' d}(\epsilon_2) \right] \cos\theta \biggr\rbrace\nonumber\\
  &&\,\times
 \lvert t_\uparrow(k_1)\, t_\downarrow(k_2)\rvert^2\, P\frac{1}{\epsilon_1-\epsilon_2},\\
   \Re &&D_0^R=-\Re D_0^A,\qquad \Im D_0^R = \Im D_0^A
 \end{eqnarray}
 \end{subequations}
(it is easy to check that in the low temperature limit, we may integrate it analytically).

\subsection{q-q two fermionic propagator}

\label{sec:q-q_two_fermionic_propagator}

Reproducing the steps  analogous to the previous case, we get
\begin{eqnarray}
 S&&_{\text{q-q}} =
 \frac{i}{4}\, \tr\left[  \check{G}_0\, \check{A}^{\text{q}} \,   \check{G}_0\, \check{A}^{\text{q}}\right]\nonumber\\
 &&=
 i\, \int dt_1\, \int dt_2 \sum_{a b} b^{\text{q}}_a(t_1)\, D_{a b}(t_1-t_2)\, b^{\text{q}}_b(t_2)\nonumber\\
\end{eqnarray}
where
\begin{eqnarray}
 D&&_{ab}(t_1-t_2) \nonumber\\
 &&=\sum_{\substack{d,s_1',s_2'\\\lvert k_1\rvert,\lvert k_2\rvert}} S(t_1-t_2)\, \mathcal{M}_{a \lvert k_1\rvert \lvert k_2\rvert}^{s_1' s_2'}(\perp)\, \mathcal{M}_{b \lvert k_2\rvert \lvert k_1\rvert}^{s_2' s_1'}(\perp)\nonumber\\
\end{eqnarray}
and
\begin{eqnarray}
 S&&(t) :=
 \frac{1}{4} \Bigl[ K^{s_1' d}_{\lvert k_1\rvert} (-t)\, K^{s_2' d}_{\lvert k_2\rvert} (t) +
 R_{\lvert k_1\rvert} (-t)\, A_{\lvert k_2\rvert} (t) \nonumber\\
 &&\quad + A_{\lvert k_1\rvert} (-t)\, R_{\lvert k_2\rvert} (t) \Bigr]\nonumber\\
 &&=
 \frac{1}{4} \biggl\{ K^{s_1' d}_{\lvert k_1\rvert} (-t)\, K^{s_2' d}_{\lvert k_2\rvert} (t)-
 \left[ R_{\lvert k_1\rvert} (-t)- A_{\lvert k_1\rvert} (-t)\right]\nonumber\\
 &&\quad\times \left[ R_{\lvert k_2\rvert} (t)- A_{\lvert k_2\rvert} (t)  \right]  \biggr\}\nonumber\\
 &&=
  \frac{1}{4} \Biggl[ K^{s_1' d}_{\lvert k_1\rvert} (-t)\, K^{s_2' d}_{\lvert k_2\rvert} (t) \nonumber\\
  &&\quad-\frac{K^{s_1' d}_{\lvert k_1\rvert} (-t)}{1-2\, n_F^{s_1' d}(\epsilon_{\lvert k_1\rvert})}\, \frac{K^{s_2' d}_{\lvert k_2\rvert} (t)}{1-2\, n_F^{s_2' d}(\epsilon_{\lvert k_2\rvert})}  \Biggr]\nonumber
 \\
  &&\xrightarrow{\text{F.T.}} \frac{1}{4}\, \int \frac{d\epsilon}{2\,\pi}\, \biggl[ K^{s_1' d}_{\lvert k_1\rvert} (\epsilon)\nonumber\\
  &&\quad\times K^{s_2' d}_{\lvert k_2\rvert} (\epsilon+\omega) 
  -\frac{K^{s_1' d}_{\lvert k_1\rvert} (\epsilon)}{1-2\, n_F^{s_1' d}(\epsilon_{\lvert k_1\rvert})}\, \frac{K^{s_2' d}_{\lvert k_2\rvert} (\epsilon+\omega)}{1-2\, n_F^{s_2' d}(\epsilon_{\lvert k_2\rvert})}  \biggr]\nonumber\\
  &&=
 \frac{\pi}{2}\, \delta(\epsilon_{\lvert k_1\rvert}-\epsilon_{\lvert k_2\rvert}+\omega)  \nonumber\\
 &&\quad\times \left\lbrace 1-\left[ 1-2\, n_F^{s_1' d}(\epsilon_{\lvert k_1\rvert})\right] \left[ 1-2\, n_F^{s_2' d}(\epsilon_{\lvert k_2\rvert})\right] \right\rbrace .\nonumber\\
\end{eqnarray}
We may simplify the expression by observing that:
\begin{subequations}
\begin{eqnarray}
 &&1-2\, n_F^{\mu}(\epsilon)=\tanh\frac{\beta\, (\epsilon-\mu)}{2},\\
 &&1-\tanh x \, \tanh y=\coth(x-y) \left[ \tanh x -\tanh y\right] ,\nonumber\\
\end{eqnarray}
\end{subequations}
from which
\begin{eqnarray}
 S&&(\omega)\nonumber\\
 &&=
 \pi\, \delta(\epsilon_{\lvert k_1\rvert}-\epsilon_{\lvert k_2\rvert}+\omega)
 \coth\frac{\beta \, ( \mu^{s_1' d}-\mu^{s_2' d} + \omega ) }{2}\nonumber\\
 &&\times
 \left[ n_F^{s_1' d}(\epsilon_{\lvert k_1\rvert}) - n_F^{s_2' d}(\epsilon_{\lvert k_2\rvert}) \right] ;
\end{eqnarray}
in particular, $S_{\text{q-q}}$ is given by the sum of three terms:
\begin{itemize}
 \item a term $i \int dt_1 \int dt_2\,  b^\text{q}(t_1)\, \tilde{D}_+(t_1-t_2)\, b^\text{q}(t_2)$, where:
 \begin{eqnarray}
  \tilde{D}&&_+(\omega)=
  \pi\, \frac{\lambda^2\, J\, N^4}{2} 
  \sum_{\substack{d, s_1', \lvert k_1\rvert, \\ s_2', \lvert k_2\rvert}}
  t^\ast_\downarrow(k_1)\, t^\ast_\downarrow(k_2)\nonumber\\
  &&\times t_\uparrow(k_1)\, t_\uparrow(k_2)\,
  \delta(\epsilon_{\lvert k_1\rvert}-\epsilon_{\lvert k_2\rvert}+\omega)\nonumber\\
  &&\times
  \coth\frac{\beta \, ( \mu^{s_1' d}-\mu^{s_2' d} + \omega ) }{2}\nonumber\\
  &&\times
 \left[ n_F^{s_1' d}(\epsilon_{\lvert k_1\rvert}) - n_F^{s_2' d}(\epsilon_{\lvert k_2\rvert}) \right]
 \mathcal{L}^{s_1' s_2'}\, \mathcal{L}^{s_2' s_1'};
 \end{eqnarray}
 \item a term $i \int dt_1 \int dt_2\,  \bar{b}^\text{q}(t_1)\, \tilde{D}_-(t_1-t_2)\, \bar{b}^\text{q}(t_2)$, where $\tilde{D}_-(\omega)$ is the complex conjugate of $\tilde{D}_-(\omega)$:
 \begin{equation}
  \tilde{D}_-(\omega)=
  \tilde{D}_+^\ast(\omega)
 \end{equation}
  \item a term $i \int dt_1 \int dt_2\,  \bar{b}^\text{q}(t_1)\, D^K(t_1-t_2)\, b^\text{q}(t_2)$, where:
 \begin{eqnarray}
  D&&^K(\omega)\nonumber\\
  &&=
  \pi\, \frac{\lambda^2\, J\, N^4}{2} 
  \sum_{\substack{d, s_1', \lvert k_1\rvert, \\ s_2', \lvert k_2\rvert}}
 \left[ n_F^{s_1' d}(\epsilon_{\lvert k_1\rvert}) - n_F^{s_2' d}(\epsilon_{\lvert k_2\rvert}) \right]\nonumber\\
 &&\,\times
 \Bigl[ \coth\frac{\beta \, ( \mu^{s_1' d}-\mu^{s_2' d} + \omega ) }{2}\, \delta(\epsilon_{\lvert k_1\rvert}-\epsilon_{\lvert k_2\rvert}+\omega)\nonumber\\
 &&\, \times
 \lvert t_\uparrow(k_1)\, t_\downarrow(k_2) \rvert^2\, (\mathcal{L}^\dagger)^{s_1' s_2'}\, \mathcal{L}^{s_2' s_1'}\nonumber\\
 &&\, +
 \coth\frac{\beta \, ( \mu^{s_1' d}-\mu^{s_2' d} - \omega ) }{2}\, \delta(\epsilon_{\lvert k_1\rvert}-\epsilon_{\lvert k_2\rvert}-\omega)\nonumber\\
 &&\, \times
 \lvert t_\downarrow(k_1)\, t_\uparrow(k_2)\rvert^2\, \mathcal{L}^{s_1' s_2'}\, (\mathcal{L}^\dagger)^{s_2' s_1'}\Bigr] .
 \end{eqnarray}
\end{itemize}

It is interesting to observe that, if we consider the equilibrium limit, that is $\mu^{s' d}=\epsilon_F$, since $\mathcal{L}^2(\perp)=0$ and $\tr[L^\dagger\, L]=1$, we have (see relations~\eqref{def:DAomega} and~\eqref{def:DRomega})
the fluctuation-dissipation theorem:
\begin{gather*}
 D^K_\text{eq}(\omega) = \coth\frac{\beta \, \omega  }{2} \left[ D^R_\text{eq}(\omega) - D^A_\text{eq}(\omega) \right] .
\end{gather*}

Now we consider again the limit $\omega\to 0$:
\begin{eqnarray}
 S&&(t_1-t_2) =
 \frac{\pi}{2}\, \delta(t_1-t_2)\, \delta(\epsilon_{\lvert k_1\rvert}-\epsilon_{\lvert k_2\rvert}) \nonumber\\
 &&\times
 \left\lbrace 1-\left[ 1-2\, n_F^{s_1' d}(\epsilon_{\lvert k_1\rvert})\right] \left[ 1-2\, n_F^{s_2' d}(\epsilon_{\lvert k_2\rvert})\right] \right\rbrace\nonumber\\
\end{eqnarray}
and then for $D_{ab}$ we have
\begin{itemize}
 \item the term proportional to $b^{\text{q}}\, b^{\text{q}}$:
 \begin{eqnarray}
  \tilde{D}&&_+(t_1-t_2)\nonumber\\
  &&=
  \delta(t_1-t_2) \sum_{\substack{d,s_1',s_2'\\ \lvert k_1\rvert,\lvert k_2\rvert}} \frac{\pi}{2}\,  \delta(\epsilon_{\lvert k_1\rvert}-\epsilon_{\lvert k_2\rvert})  \nonumber\\
  &&\quad \times
  \left\lbrace 1-\left[ 1-2\, n_F^{s_1' d}(\epsilon_{\lvert k_1\rvert})\right] \left[ 1-2\, n_F^{s_2' d}(\epsilon_{\lvert k_2\rvert})\right] \right\rbrace\nonumber\\
  &&\quad \times
  \mathcal{M}^{s_1' s_2'}_{\lvert k_1\rvert \lvert k_2\rvert}(\perp)\,  \mathcal{M}^{s_2' s_1'}_{\lvert k_2\rvert \lvert k_1\rvert}(\perp)\nonumber\\
  &&=
  -\delta(t_1-t_2)\, \frac{\pi\, m\, \lambda^2\, J\, N^4}{8}\, e^{-2\, i\, \phi}\, \sin^2\theta  \nonumber\\
  &&\quad \times
  \sum_d \int_0^\infty \frac{d\epsilon}{\epsilon} \left[ t_\downarrow^\ast(k) \,   t_\uparrow(k) \right]^2  \left[ n_F^{\uparrow' d}(\epsilon)-n_F^{\downarrow' d}(\epsilon) \right]^2\nonumber\\
 \end{eqnarray}
and in the low temperature and differential potentials limit
\begin{eqnarray}
  &&\left[ n_F^{\uparrow' d}(\epsilon)-n_F^{\downarrow' d}(\epsilon) \right]^2 \nonumber\\
  &&\quad= \left[ n_F^{\uparrow' d}(\epsilon)-n_F^{\downarrow' d}(\epsilon) \right] \sign(\mu^{\uparrow' d}-\mu^{\downarrow' d})\nonumber\\
   &&\quad\sim
   \delta(\epsilon-\epsilon_F) \lvert \mu^{\uparrow' d}-\mu^{\downarrow' d} \rvert;
\end{eqnarray}
\item the term proportional to $\bar{b}^{\text{q}}\, \bar{b}^{\text{q}}$, that is $ \tilde{D}_-(t_1-t_2)= \tilde{D}_+^\ast(t_1-t_2)$;
\item the term proportional to $\bar{b}^{\text{q}}\, b^{\text{q}}$:
\begin{eqnarray}
 D&&^K(t_1-t_2) \nonumber\\
 &&=
 2\, \delta(t_1-t_2) \sum_{\substack{d,s_1',s_2'\\ \lvert k_1\rvert,\lvert k_2\rvert}} \frac{\pi}{2}\,  \delta(\epsilon_{\lvert k_1\rvert}-\epsilon_{\lvert k_2\rvert})  \nonumber\\
 &&\,\times
  \left\lbrace 1-\left[ 1-2\, n_F^{s_1' d}(\epsilon_{\lvert k_1\rvert})\right] \left[ 1-2\, n_F^{s_2' d}(\epsilon_{\lvert k_2\rvert})\right] \right\rbrace\nonumber\\
  &&\, \times
  \left( \mathcal{M}^\dagger\right) ^{s_1' s_2'}_{\lvert k_1\rvert \lvert k_2\rvert}(\perp)\, \mathcal{M}^{s_2' s_1'}_{\lvert k_2\rvert \lvert k_1\rvert}(\perp)\nonumber\\
  &&=
  \delta(t_1-t_2)\, \frac{\pi\, m\, \lambda^2\, J\, N^4}{8}  
  \sum_d \int_0^\infty \frac{d\epsilon}{\epsilon} \,\lvert t_\downarrow(k)  \rvert^2 \nonumber\\
  &&\, \times\lvert   t_\uparrow(k) \rvert^2 
   \biggl\lbrace \left[ n_F^{\uparrow' d}(\epsilon)-n_F^{\downarrow' d}(\epsilon) \right]^2 
 \left( -2\, \sin^2\theta \right) \nonumber\\
 &&\, +
  4 \left[ n_F^{\uparrow' d}(\epsilon)+n_F^{\downarrow' d}(\epsilon) -2\,  n_F^{\uparrow' d}(\epsilon)\, n_F^{\downarrow' d}(\epsilon)  \right]  \biggr\rbrace 
\end{eqnarray}
and in the low temperature and differential potentials limit, where $\left( n_F^{s' d}\right) ^2 = n_F^{s' d}$:
\begin{eqnarray}
 &&\left[ n_F^{\uparrow' d}(\epsilon)-n_F^{\downarrow' d}(\epsilon) \right]^2 
 \left( -2\, \sin^2\theta \right) \nonumber\\
 &&\quad+
  4 \left[ n_F^{\uparrow' d}(\epsilon)+n_F^{\downarrow' d}(\epsilon) -2\,  n_F^{\uparrow' d}(\epsilon)\, n_F^{\downarrow' d}(\epsilon)  \right] \nonumber\\
 && \, =
  2 \left[ n_F^{\uparrow' d}(\epsilon)-n_F^{\downarrow' d}(\epsilon) \right]^2 
 \left( 2-\sin^2\theta \right)\nonumber\\
 &&\, =
  2\, \lvert \mu^{\uparrow' d}-\mu^{\downarrow' d} \rvert \left( 2-\sin^2\theta \right)  \delta(\epsilon-\epsilon_F).
\end{eqnarray}

\end{itemize}

\section{Fokker-Planck equation}

\label{sec:Fokker-Planck_equation}

Here we briefly review the relation between Langevin and Fokker-Planck equations~\cite{ottinger1996stochastic,chandrasekhar1943stochastic}.  If we have a stochastic differential equation of the form
\begin{equation}
 d\pmb{X}_t= \pmb{\mu}(\pmb{X}_t,t)\, dt + \pmb{\sigma}(\pmb{X}_t,t)\, d\pmb{W}_t,
 \label{eqn:generalized_Brown}
\end{equation}
where $\pmb{X}_t$ is an $N$-dimensional column vector of unknown functions, $\pmb{W}_t$ is an $M$-dimensional column vector of independent standard Wiener processes, $\pmb{\mu}$  is called \emph{drift vector}, $\pmb{\sigma}$ is an $N\times M$-dimensional matrix and
\begin{equation}
 \pmb{D}:=\frac{1}{2}\, \pmb{\sigma}\, \pmb{\sigma}^t
\end{equation}
is called \emph{diffusion tensor}, we have that Eq.~\eqref{eqn:generalized_Brown} is equivalent to the probability density equation (Fokker-Planck equation):
\begin{eqnarray}
 \frac{\partial p(\pmb{x},t)}{\partial t}&=&-\sum_{i=1}^N \frac{\partial}{\partial x_i}\left[ \mu_i(\pmb{x},t)\, p(\pmb{x},t)\right] \nonumber\\
 &&+
 \sum_{i,j=1}^N  \frac{\partial^2}{\partial x_i\, \partial x_j}  \left[ D_{ij}(\pmb{x},t)\, p(\pmb{x},t)\right],
 \label{eqn:genelized_Fokker-Planck}
\end{eqnarray}
if the It\={o} regularization is assumed.


\normalem


\bibliography{Bibliography}

\begin{thebibliography}{47}%
\makeatletter
\providecommand \@ifxundefined [1]{%
 \@ifx{#1\undefined}
}%
\providecommand \@ifnum [1]{%
 \ifnum #1\expandafter \@firstoftwo
 \else \expandafter \@secondoftwo
 \fi
}%
\providecommand \@ifx [1]{%
 \ifx #1\expandafter \@firstoftwo
 \else \expandafter \@secondoftwo
 \fi
}%
\providecommand \natexlab [1]{#1}%
\providecommand \enquote  [1]{``#1''}%
\providecommand \bibnamefont  [1]{#1}%
\providecommand \bibfnamefont [1]{#1}%
\providecommand \citenamefont [1]{#1}%
\providecommand \href@noop [0]{\@secondoftwo}%
\providecommand \href [0]{\begingroup \@sanitize@url \@href}%
\providecommand \@href[1]{\@@startlink{#1}\@@href}%
\providecommand \@@href[1]{\endgroup#1\@@endlink}%
\providecommand \@sanitize@url [0]{\catcode `\\12\catcode `\$12\catcode
  `\&12\catcode `\#12\catcode `\^12\catcode `\_12\catcode `\%12\relax}%
\providecommand \@@startlink[1]{}%
\providecommand \@@endlink[0]{}%
\providecommand \url  [0]{\begingroup\@sanitize@url \@url }%
\providecommand \@url [1]{\endgroup\@href {#1}{\urlprefix }}%
\providecommand \urlprefix  [0]{URL }%
\providecommand \Eprint [0]{\href }%
\providecommand \doibase [0]{https://doi.org/}%
\providecommand \selectlanguage [0]{\@gobble}%
\providecommand \bibinfo  [0]{\@secondoftwo}%
\providecommand \bibfield  [0]{\@secondoftwo}%
\providecommand \translation [1]{[#1]}%
\providecommand \BibitemOpen [0]{}%
\providecommand \bibitemStop [0]{}%
\providecommand \bibitemNoStop [0]{.\EOS\space}%
\providecommand \EOS [0]{\spacefactor3000\relax}%
\providecommand \BibitemShut  [1]{\csname bibitem#1\endcsname}%
\let\auto@bib@innerbib\@empty
\bibitem [{\citenamefont {Pesin}\ and\ \citenamefont
  {MacDonald}(2012)}]{pesin2012spintronics}%
  \BibitemOpen
  \bibfield  {author} {\bibinfo {author} {\bibfnamefont {D.}~\bibnamefont
  {Pesin}}\ and\ \bibinfo {author} {\bibfnamefont {A.~H.}\ \bibnamefont
  {MacDonald}},\ }\bibfield  {title} {\bibinfo {title} {Spintronics and
  pseudospintronics in graphene and topological insulators},\ }\href@noop {}
  {\bibfield  {journal} {\bibinfo  {journal} {Nature Materials}\ }\textbf
  {\bibinfo {volume} {11}},\ \bibinfo {pages} {409} (\bibinfo {year}
  {2012})}\BibitemShut {NoStop}%
\bibitem [{\citenamefont
  {Slonczewski}(1996)}]{Slonczewski:Current-drivenExcitation}%
  \BibitemOpen
  \bibfield  {author} {\bibinfo {author} {\bibfnamefont {J.}~\bibnamefont
  {Slonczewski}},\ }\bibfield  {title} {\bibinfo {title} {Current-driven
  excitation of magnetic multilayers},\ }\href@noop {} {\bibfield  {journal}
  {\bibinfo  {journal} {Journal of Magnetism and Magnetic Materials}\ }\textbf
  {\bibinfo {volume} {159}},\ \bibinfo {pages} {L1} (\bibinfo {year}
  {1996})}\BibitemShut {NoStop}%
\bibitem [{\citenamefont {Berger}(1996)}]{Berger1996}%
  \BibitemOpen
  \bibfield  {author} {\bibinfo {author} {\bibfnamefont {L.}~\bibnamefont
  {Berger}},\ }\bibfield  {title} {\bibinfo {title} {Emission of spin waves by
  a magnetic multilayer traversed by a current},\ }\href@noop {} {\bibfield
  {journal} {\bibinfo  {journal} {Physical Review B}\ }\textbf {\bibinfo
  {volume} {54}},\ \bibinfo {pages} {9353} (\bibinfo {year}
  {1996})}\BibitemShut {NoStop}%
\bibitem [{\citenamefont {Landau}\ and\ \citenamefont
  {Lifshitz}(1935)}]{landau1935theory}%
  \BibitemOpen
  \bibfield  {author} {\bibinfo {author} {\bibfnamefont {L.~D.}\ \bibnamefont
  {Landau}}\ and\ \bibinfo {author} {\bibfnamefont {E.}~\bibnamefont
  {Lifshitz}},\ }\bibfield  {title} {\bibinfo {title} {On the theory of the
  dispersion of magnetic permeability in ferromagnetic bodies},\ }\href@noop {}
  {\bibfield  {journal} {\bibinfo  {journal} {Phys. Z. Sowjetunion}\ }\textbf
  {\bibinfo {volume} {8}},\ \bibinfo {pages} {101} (\bibinfo {year}
  {1935})}\BibitemShut {NoStop}%
\bibitem [{\citenamefont {Gilbert}(2004)}]{gilbert2004phenomenological}%
  \BibitemOpen
  \bibfield  {author} {\bibinfo {author} {\bibfnamefont {T.~L.}\ \bibnamefont
  {Gilbert}},\ }\bibfield  {title} {\bibinfo {title} {A phenomenological theory
  of damping in ferromagnetic materials},\ }\href@noop {} {\bibfield  {journal}
  {\bibinfo  {journal} {IEEE Transactions on Magnetics}\ }\textbf {\bibinfo
  {volume} {40}},\ \bibinfo {pages} {3443} (\bibinfo {year}
  {2004})}\BibitemShut {NoStop}%
\bibitem [{\citenamefont {Xia}\ \emph {et~al.}(2002)\citenamefont {Xia},
  \citenamefont {Kelly}, \citenamefont {Bauer}, \citenamefont {Brataas},\ and\
  \citenamefont {Turek}}]{xia2002spin}%
  \BibitemOpen
  \bibfield  {author} {\bibinfo {author} {\bibfnamefont {K.}~\bibnamefont
  {Xia}}, \bibinfo {author} {\bibfnamefont {P.~J.}\ \bibnamefont {Kelly}},
  \bibinfo {author} {\bibfnamefont {G.}~\bibnamefont {Bauer}}, \bibinfo
  {author} {\bibfnamefont {A.}~\bibnamefont {Brataas}},\ and\ \bibinfo {author}
  {\bibfnamefont {I.}~\bibnamefont {Turek}},\ }\bibfield  {title} {\bibinfo
  {title} {Spin torques in ferromagnetic/normal-metal structures},\ }\href@noop
  {} {\bibfield  {journal} {\bibinfo  {journal} {Physical Review B}\ }\textbf
  {\bibinfo {volume} {65}},\ \bibinfo {pages} {220401} (\bibinfo {year}
  {2002})}\BibitemShut {NoStop}%
\bibitem [{\citenamefont {Brataas}\ \emph {et~al.}(2001)\citenamefont
  {Brataas}, \citenamefont {Nazarov},\ and\ \citenamefont
  {Bauer}}]{brataas2001spin}%
  \BibitemOpen
  \bibfield  {author} {\bibinfo {author} {\bibfnamefont {A.}~\bibnamefont
  {Brataas}}, \bibinfo {author} {\bibfnamefont {Y.~V.}\ \bibnamefont
  {Nazarov}},\ and\ \bibinfo {author} {\bibfnamefont {G.~E.}\ \bibnamefont
  {Bauer}},\ }\bibfield  {title} {\bibinfo {title} {Spin-transport in
  multi-terminal normal metal-ferromagnet systems with non-collinear
  magnetizations},\ }\href@noop {} {\bibfield  {journal} {\bibinfo  {journal}
  {The European Physical Journal B-Condensed Matter and Complex Systems}\
  }\textbf {\bibinfo {volume} {22}},\ \bibinfo {pages} {99} (\bibinfo {year}
  {2001})}\BibitemShut {NoStop}%
\bibitem [{\citenamefont {Stiles}\ and\ \citenamefont
  {Zangwill}(2002)}]{stiles2002anatomy}%
  \BibitemOpen
  \bibfield  {author} {\bibinfo {author} {\bibfnamefont {M.~D.}\ \bibnamefont
  {Stiles}}\ and\ \bibinfo {author} {\bibfnamefont {A.}~\bibnamefont
  {Zangwill}},\ }\bibfield  {title} {\bibinfo {title} {Anatomy of spin-transfer
  torque},\ }\href@noop {} {\bibfield  {journal} {\bibinfo  {journal} {Physical
  Review B}\ }\textbf {\bibinfo {volume} {66}},\ \bibinfo {pages} {014407}
  (\bibinfo {year} {2002})}\BibitemShut {NoStop}%
\bibitem [{\citenamefont {Hankiewicz}\ \emph {et~al.}(2007)\citenamefont
  {Hankiewicz}, \citenamefont {Vignale},\ and\ \citenamefont
  {Tserkovnyak}}]{hankiewicz2007gilbert}%
  \BibitemOpen
  \bibfield  {author} {\bibinfo {author} {\bibfnamefont {E.~M.}\ \bibnamefont
  {Hankiewicz}}, \bibinfo {author} {\bibfnamefont {G.}~\bibnamefont
  {Vignale}},\ and\ \bibinfo {author} {\bibfnamefont {Y.}~\bibnamefont
  {Tserkovnyak}},\ }\bibfield  {title} {\bibinfo {title} {Gilbert damping and
  spin coulomb drag in a magnetized electron liquid with spin-orbit
  interaction},\ }\href@noop {} {\bibfield  {journal} {\bibinfo  {journal}
  {Physical Review B}\ }\textbf {\bibinfo {volume} {75}},\ \bibinfo {pages}
  {174434} (\bibinfo {year} {2007})}\BibitemShut {NoStop}%
\bibitem [{\citenamefont {Ralph}\ and\ \citenamefont
  {Stiles}(2008)}]{ralph2008spin}%
  \BibitemOpen
  \bibfield  {author} {\bibinfo {author} {\bibfnamefont {D.~C.}\ \bibnamefont
  {Ralph}}\ and\ \bibinfo {author} {\bibfnamefont {M.~D.}\ \bibnamefont
  {Stiles}},\ }\bibfield  {title} {\bibinfo {title} {Spin transfer torques},\
  }\href@noop {} {\bibfield  {journal} {\bibinfo  {journal} {Journal of
  Magnetism and Magnetic Materials}\ }\textbf {\bibinfo {volume} {320}},\
  \bibinfo {pages} {1190} (\bibinfo {year} {2008})}\BibitemShut {NoStop}%
\bibitem [{\citenamefont {Hankiewicz}\ \emph {et~al.}(2008)\citenamefont
  {Hankiewicz}, \citenamefont {Vignale},\ and\ \citenamefont
  {Tserkovnyak}}]{hankiewicz2008inhomogeneous}%
  \BibitemOpen
  \bibfield  {author} {\bibinfo {author} {\bibfnamefont {E.~M.}\ \bibnamefont
  {Hankiewicz}}, \bibinfo {author} {\bibfnamefont {G.}~\bibnamefont
  {Vignale}},\ and\ \bibinfo {author} {\bibfnamefont {Y.}~\bibnamefont
  {Tserkovnyak}},\ }\bibfield  {title} {\bibinfo {title} {Inhomogeneous gilbert
  damping from impurities and electron-electron interactions},\ }\href@noop {}
  {\bibfield  {journal} {\bibinfo  {journal} {Physical Review B}\ }\textbf
  {\bibinfo {volume} {78}},\ \bibinfo {pages} {020404} (\bibinfo {year}
  {2008})}\BibitemShut {NoStop}%
\bibitem [{\citenamefont {Tatara}\ \emph {et~al.}(2008)\citenamefont {Tatara},
  \citenamefont {Kohno},\ and\ \citenamefont
  {Shibata}}]{tatara2008microscopic}%
  \BibitemOpen
  \bibfield  {author} {\bibinfo {author} {\bibfnamefont {G.}~\bibnamefont
  {Tatara}}, \bibinfo {author} {\bibfnamefont {H.}~\bibnamefont {Kohno}},\ and\
  \bibinfo {author} {\bibfnamefont {J.}~\bibnamefont {Shibata}},\ }\bibfield
  {title} {\bibinfo {title} {Microscopic approach to current-driven domain wall
  dynamics},\ }\href@noop {} {\bibfield  {journal} {\bibinfo  {journal}
  {Physics Reports}\ }\textbf {\bibinfo {volume} {468}},\ \bibinfo {pages}
  {213} (\bibinfo {year} {2008})}\BibitemShut {NoStop}%
\bibitem [{\citenamefont {Tserkovnyak}\ \emph {et~al.}(2009)\citenamefont
  {Tserkovnyak}, \citenamefont {Hankiewicz},\ and\ \citenamefont
  {Vignale}}]{tserkovnyak2009transverse}%
  \BibitemOpen
  \bibfield  {author} {\bibinfo {author} {\bibfnamefont {Y.}~\bibnamefont
  {Tserkovnyak}}, \bibinfo {author} {\bibfnamefont {E.~M.}\ \bibnamefont
  {Hankiewicz}},\ and\ \bibinfo {author} {\bibfnamefont {G.}~\bibnamefont
  {Vignale}},\ }\bibfield  {title} {\bibinfo {title} {Transverse spin diffusion
  in ferromagnets},\ }\href@noop {} {\bibfield  {journal} {\bibinfo  {journal}
  {Physical Review B}\ }\textbf {\bibinfo {volume} {79}},\ \bibinfo {pages}
  {094415} (\bibinfo {year} {2009})}\BibitemShut {NoStop}%
\bibitem [{\citenamefont {Garate}\ \emph {et~al.}(2009)\citenamefont {Garate},
  \citenamefont {Gilmore}, \citenamefont {Stiles},\ and\ \citenamefont
  {MacDonald}}]{garate2009nonadiabatic}%
  \BibitemOpen
  \bibfield  {author} {\bibinfo {author} {\bibfnamefont {I.}~\bibnamefont
  {Garate}}, \bibinfo {author} {\bibfnamefont {K.}~\bibnamefont {Gilmore}},
  \bibinfo {author} {\bibfnamefont {M.~D.}\ \bibnamefont {Stiles}},\ and\
  \bibinfo {author} {\bibfnamefont {A.~H.}\ \bibnamefont {MacDonald}},\
  }\bibfield  {title} {\bibinfo {title} {Nonadiabatic spin-transfer torque in
  real materials},\ }\href@noop {} {\bibfield  {journal} {\bibinfo  {journal}
  {Physical Review B}\ }\textbf {\bibinfo {volume} {79}},\ \bibinfo {pages}
  {104416} (\bibinfo {year} {2009})}\BibitemShut {NoStop}%
\bibitem [{\citenamefont {Tatara}(2018)}]{tatara2018effective}%
  \BibitemOpen
  \bibfield  {author} {\bibinfo {author} {\bibfnamefont {G.}~\bibnamefont
  {Tatara}},\ }\bibfield  {title} {\bibinfo {title} {Effective gauge field
  theory of spintronics},\ }\href@noop {} {\bibfield  {journal} {\bibinfo
  {journal} {Physica E: Low-dimensional Systems and Nanostructures}\ }
  (\bibinfo {year} {2018})}\BibitemShut {NoStop}%
\bibitem [{\citenamefont {Tserkovnyak}\ \emph {et~al.}(2005)\citenamefont
  {Tserkovnyak}, \citenamefont {Brataas}, \citenamefont {Bauer},\ and\
  \citenamefont {Halperin}}]{Tserkovnyak2005}%
  \BibitemOpen
  \bibfield  {author} {\bibinfo {author} {\bibfnamefont {Y.}~\bibnamefont
  {Tserkovnyak}}, \bibinfo {author} {\bibfnamefont {A.}~\bibnamefont
  {Brataas}}, \bibinfo {author} {\bibfnamefont {G.~E.~W.}\ \bibnamefont
  {Bauer}},\ and\ \bibinfo {author} {\bibfnamefont {B.~I.}\ \bibnamefont
  {Halperin}},\ }\bibfield  {title} {\bibinfo {title} {Nonlocal magnetization
  dynamics in ferromagnetic heterostructures},\ }\href
  {https://doi.org/10.1103/RevModPhys.77.1375} {\bibfield  {journal} {\bibinfo
  {journal} {Reviews of Modern Physics}\ }\textbf {\bibinfo {volume} {77}},\
  \bibinfo {pages} {1375} (\bibinfo {year} {2005})}\BibitemShut {NoStop}%
\bibitem [{\citenamefont {Brataas}\ \emph {et~al.}(2006)\citenamefont
  {Brataas}, \citenamefont {Bauer},\ and\ \citenamefont {Kelly}}]{BRATAAS2006}%
  \BibitemOpen
  \bibfield  {author} {\bibinfo {author} {\bibfnamefont {A.}~\bibnamefont
  {Brataas}}, \bibinfo {author} {\bibfnamefont {G.~E.}\ \bibnamefont {Bauer}},\
  and\ \bibinfo {author} {\bibfnamefont {P.~J.}\ \bibnamefont {Kelly}},\
  }\bibfield  {title} {\bibinfo {title} {Non-collinear magnetoelectronics},\
  }\href@noop {} {\bibfield  {journal} {\bibinfo  {journal} {Physics Reports}\
  }\textbf {\bibinfo {volume} {427}},\ \bibinfo {pages} {157 } (\bibinfo {year}
  {2006})}\BibitemShut {NoStop}%
\bibitem [{\citenamefont {Hellman}\ \emph {et~al.}(2017)\citenamefont
  {Hellman}, \citenamefont {Hoffmann}, \citenamefont {Tserkovnyak},
  \citenamefont {Beach}, \citenamefont {Fullerton}, \citenamefont {Leighton},
  \citenamefont {MacDonald}, \citenamefont {Ralph}, \citenamefont {Arena},
  \citenamefont {D\"urr}, \citenamefont {Fischer}, \citenamefont {Grollier},
  \citenamefont {Heremans}, \citenamefont {Jungwirth}, \citenamefont {Kimel},
  \citenamefont {Koopmans}, \citenamefont {Krivorotov}, \citenamefont {May},
  \citenamefont {Petford-Long}, \citenamefont {Rondinelli}, \citenamefont
  {Samarth}, \citenamefont {Schuller}, \citenamefont {Slavin}, \citenamefont
  {Stiles}, \citenamefont {Tchernyshyov}, \citenamefont {Thiaville},\ and\
  \citenamefont {Zink}}]{Hellman2017}%
  \BibitemOpen
  \bibfield  {author} {\bibinfo {author} {\bibfnamefont {F.}~\bibnamefont
  {Hellman}}, \bibinfo {author} {\bibfnamefont {A.}~\bibnamefont {Hoffmann}},
  \bibinfo {author} {\bibfnamefont {Y.}~\bibnamefont {Tserkovnyak}}, \bibinfo
  {author} {\bibfnamefont {G.~S.~D.}\ \bibnamefont {Beach}}, \bibinfo {author}
  {\bibfnamefont {E.~E.}\ \bibnamefont {Fullerton}}, \bibinfo {author}
  {\bibfnamefont {C.}~\bibnamefont {Leighton}}, \bibinfo {author}
  {\bibfnamefont {A.~H.}\ \bibnamefont {MacDonald}}, \bibinfo {author}
  {\bibfnamefont {D.~C.}\ \bibnamefont {Ralph}}, \bibinfo {author}
  {\bibfnamefont {D.~A.}\ \bibnamefont {Arena}}, \bibinfo {author}
  {\bibfnamefont {H.~A.}\ \bibnamefont {D\"urr}}, \bibinfo {author}
  {\bibfnamefont {P.}~\bibnamefont {Fischer}}, \bibinfo {author} {\bibfnamefont
  {J.}~\bibnamefont {Grollier}}, \bibinfo {author} {\bibfnamefont {J.~P.}\
  \bibnamefont {Heremans}}, \bibinfo {author} {\bibfnamefont {T.}~\bibnamefont
  {Jungwirth}}, \bibinfo {author} {\bibfnamefont {A.~V.}\ \bibnamefont
  {Kimel}}, \bibinfo {author} {\bibfnamefont {B.}~\bibnamefont {Koopmans}},
  \bibinfo {author} {\bibfnamefont {I.~N.}\ \bibnamefont {Krivorotov}},
  \bibinfo {author} {\bibfnamefont {S.~J.}\ \bibnamefont {May}}, \bibinfo
  {author} {\bibfnamefont {A.~K.}\ \bibnamefont {Petford-Long}}, \bibinfo
  {author} {\bibfnamefont {J.~M.}\ \bibnamefont {Rondinelli}}, \bibinfo
  {author} {\bibfnamefont {N.}~\bibnamefont {Samarth}}, \bibinfo {author}
  {\bibfnamefont {I.~K.}\ \bibnamefont {Schuller}}, \bibinfo {author}
  {\bibfnamefont {A.~N.}\ \bibnamefont {Slavin}}, \bibinfo {author}
  {\bibfnamefont {M.~D.}\ \bibnamefont {Stiles}}, \bibinfo {author}
  {\bibfnamefont {O.}~\bibnamefont {Tchernyshyov}}, \bibinfo {author}
  {\bibfnamefont {A.}~\bibnamefont {Thiaville}},\ and\ \bibinfo {author}
  {\bibfnamefont {B.~L.}\ \bibnamefont {Zink}},\ }\bibfield  {title} {\bibinfo
  {title} {Interface-induced phenomena in magnetism},\ }\href@noop {}
  {\bibfield  {journal} {\bibinfo  {journal} {Reviews of Modern Physics}\
  }\textbf {\bibinfo {volume} {89}},\ \bibinfo {pages} {025006} (\bibinfo
  {year} {2017})}\BibitemShut {NoStop}%
\bibitem [{\citenamefont {Brataas}\ \emph {et~al.}(2000)\citenamefont
  {Brataas}, \citenamefont {Nazarov},\ and\ \citenamefont
  {Bauer}}]{Braatas2000}%
  \BibitemOpen
  \bibfield  {author} {\bibinfo {author} {\bibfnamefont {A.}~\bibnamefont
  {Brataas}}, \bibinfo {author} {\bibfnamefont {Y.~V.}\ \bibnamefont
  {Nazarov}},\ and\ \bibinfo {author} {\bibfnamefont {G.~E.~W.}\ \bibnamefont
  {Bauer}},\ }\bibfield  {title} {\bibinfo {title} {Finite-element theory of
  transport in ferromagnet--normal metal systems},\ }\href@noop {} {\bibfield
  {journal} {\bibinfo  {journal} {Physical Review Letters}\ }\textbf {\bibinfo
  {volume} {84}},\ \bibinfo {pages} {2481} (\bibinfo {year}
  {2000})}\BibitemShut {NoStop}%
\bibitem [{\citenamefont {Devolder}\ \emph {et~al.}(2008)\citenamefont
  {Devolder}, \citenamefont {Hayakawa}, \citenamefont {Ito}, \citenamefont
  {Takahashi}, \citenamefont {Ikeda}, \citenamefont {Crozat}, \citenamefont
  {Zerounian}, \citenamefont {Kim}, \citenamefont {Chappert},\ and\
  \citenamefont {Ohno}}]{devolder2008single}%
  \BibitemOpen
  \bibfield  {author} {\bibinfo {author} {\bibfnamefont {T.}~\bibnamefont
  {Devolder}}, \bibinfo {author} {\bibfnamefont {J.}~\bibnamefont {Hayakawa}},
  \bibinfo {author} {\bibfnamefont {K.}~\bibnamefont {Ito}}, \bibinfo {author}
  {\bibfnamefont {H.}~\bibnamefont {Takahashi}}, \bibinfo {author}
  {\bibfnamefont {S.}~\bibnamefont {Ikeda}}, \bibinfo {author} {\bibfnamefont
  {P.}~\bibnamefont {Crozat}}, \bibinfo {author} {\bibfnamefont
  {N.}~\bibnamefont {Zerounian}}, \bibinfo {author} {\bibfnamefont {J.-V.}\
  \bibnamefont {Kim}}, \bibinfo {author} {\bibfnamefont {C.}~\bibnamefont
  {Chappert}},\ and\ \bibinfo {author} {\bibfnamefont {H.}~\bibnamefont
  {Ohno}},\ }\bibfield  {title} {\bibinfo {title} {Single-shot time-resolved
  measurements of nanosecond-scale spin-transfer induced switching: Stochastic
  versus deterministic aspects},\ }\href@noop {} {\bibfield  {journal}
  {\bibinfo  {journal} {Physical Review Letters}\ }\textbf {\bibinfo {volume}
  {100}},\ \bibinfo {pages} {057206} (\bibinfo {year} {2008})}\BibitemShut
  {NoStop}%
\bibitem [{\citenamefont {Tomita}\ \emph {et~al.}(2008)\citenamefont {Tomita},
  \citenamefont {Konishi}, \citenamefont {Nozaki}, \citenamefont {Kubota},
  \citenamefont {Fukushima}, \citenamefont {Yakushiji}, \citenamefont {Yuasa},
  \citenamefont {Nakatani}, \citenamefont {Shinjo}, \citenamefont {Shiraishi}
  \emph {et~al.}}]{tomita2008single}%
  \BibitemOpen
  \bibfield  {author} {\bibinfo {author} {\bibfnamefont {H.}~\bibnamefont
  {Tomita}}, \bibinfo {author} {\bibfnamefont {K.}~\bibnamefont {Konishi}},
  \bibinfo {author} {\bibfnamefont {T.}~\bibnamefont {Nozaki}}, \bibinfo
  {author} {\bibfnamefont {H.}~\bibnamefont {Kubota}}, \bibinfo {author}
  {\bibfnamefont {A.}~\bibnamefont {Fukushima}}, \bibinfo {author}
  {\bibfnamefont {K.}~\bibnamefont {Yakushiji}}, \bibinfo {author}
  {\bibfnamefont {S.}~\bibnamefont {Yuasa}}, \bibinfo {author} {\bibfnamefont
  {Y.}~\bibnamefont {Nakatani}}, \bibinfo {author} {\bibfnamefont
  {T.}~\bibnamefont {Shinjo}}, \bibinfo {author} {\bibfnamefont
  {M.}~\bibnamefont {Shiraishi}}, \emph {et~al.},\ }\bibfield  {title}
  {\bibinfo {title} {Single-shot measurements of spin-transfer switching in
  cofeb/mgo/cofeb magnetic tunnel junctions},\ }\href@noop {} {\bibfield
  {journal} {\bibinfo  {journal} {Applied Physics Express}\ }\textbf {\bibinfo
  {volume} {1}},\ \bibinfo {pages} {061303} (\bibinfo {year}
  {2008})}\BibitemShut {NoStop}%
\bibitem [{\citenamefont {Cui}\ \emph {et~al.}(2010)\citenamefont {Cui},
  \citenamefont {Finocchio}, \citenamefont {Wang}, \citenamefont {Katine},
  \citenamefont {Buhrman},\ and\ \citenamefont {Ralph}}]{cui2010single}%
  \BibitemOpen
  \bibfield  {author} {\bibinfo {author} {\bibfnamefont {Y.-T.}\ \bibnamefont
  {Cui}}, \bibinfo {author} {\bibfnamefont {G.}~\bibnamefont {Finocchio}},
  \bibinfo {author} {\bibfnamefont {C.}~\bibnamefont {Wang}}, \bibinfo {author}
  {\bibfnamefont {J.~A.}\ \bibnamefont {Katine}}, \bibinfo {author}
  {\bibfnamefont {R.~A.}\ \bibnamefont {Buhrman}},\ and\ \bibinfo {author}
  {\bibfnamefont {D.~C.}\ \bibnamefont {Ralph}},\ }\bibfield  {title} {\bibinfo
  {title} {Single-shot time-domain studies of spin-torque-driven switching in
  magnetic tunnel junctions},\ }\href@noop {} {\bibfield  {journal} {\bibinfo
  {journal} {Physical Review Letters}\ }\textbf {\bibinfo {volume} {104}},\
  \bibinfo {pages} {097201} (\bibinfo {year} {2010})}\BibitemShut {NoStop}%
\bibitem [{\citenamefont {Cheng}\ \emph {et~al.}(2010)\citenamefont {Cheng},
  \citenamefont {Boone}, \citenamefont {Zhu},\ and\ \citenamefont
  {Krivorotov}}]{cheng2010nonadiabatic}%
  \BibitemOpen
  \bibfield  {author} {\bibinfo {author} {\bibfnamefont {X.}~\bibnamefont
  {Cheng}}, \bibinfo {author} {\bibfnamefont {C.~T.}\ \bibnamefont {Boone}},
  \bibinfo {author} {\bibfnamefont {J.}~\bibnamefont {Zhu}},\ and\ \bibinfo
  {author} {\bibfnamefont {I.~N.}\ \bibnamefont {Krivorotov}},\ }\bibfield
  {title} {\bibinfo {title} {Nonadiabatic stochastic resonance of a nanomagnet
  excited by spin torque},\ }\href@noop {} {\bibfield  {journal} {\bibinfo
  {journal} {Physical Review Letters}\ }\textbf {\bibinfo {volume} {105}},\
  \bibinfo {pages} {047202} (\bibinfo {year} {2010})}\BibitemShut {NoStop}%
\bibitem [{\citenamefont {Ludwig}\ \emph {et~al.}(2017)\citenamefont {Ludwig},
  \citenamefont {Burmistrov}, \citenamefont {Gefen},\ and\ \citenamefont
  {Shnirman}}]{ludwig2017strong}%
  \BibitemOpen
  \bibfield  {author} {\bibinfo {author} {\bibfnamefont {T.}~\bibnamefont
  {Ludwig}}, \bibinfo {author} {\bibfnamefont {I.~S.}\ \bibnamefont
  {Burmistrov}}, \bibinfo {author} {\bibfnamefont {Y.}~\bibnamefont {Gefen}},\
  and\ \bibinfo {author} {\bibfnamefont {A.}~\bibnamefont {Shnirman}},\
  }\bibfield  {title} {\bibinfo {title} {Strong nonequilibrium effects in
  spin-torque systems},\ }\href@noop {} {\bibfield  {journal} {\bibinfo
  {journal} {Physical Review B}\ }\textbf {\bibinfo {volume} {95}},\ \bibinfo
  {pages} {075425} (\bibinfo {year} {2017})}\BibitemShut {NoStop}%
\bibitem [{\citenamefont {Swiebodzinski}\ \emph {et~al.}(2010)\citenamefont
  {Swiebodzinski}, \citenamefont {Chudnovskiy}, \citenamefont {Dunn},\ and\
  \citenamefont {Kamenev}}]{swiebodzinski2010spin}%
  \BibitemOpen
  \bibfield  {author} {\bibinfo {author} {\bibfnamefont {J.}~\bibnamefont
  {Swiebodzinski}}, \bibinfo {author} {\bibfnamefont {A.}~\bibnamefont
  {Chudnovskiy}}, \bibinfo {author} {\bibfnamefont {T.}~\bibnamefont {Dunn}},\
  and\ \bibinfo {author} {\bibfnamefont {A.}~\bibnamefont {Kamenev}},\
  }\bibfield  {title} {\bibinfo {title} {Spin torque dynamics with noise in
  magnetic nanosystems},\ }\href@noop {} {\bibfield  {journal} {\bibinfo
  {journal} {Physical Review B}\ }\textbf {\bibinfo {volume} {82}},\ \bibinfo
  {pages} {144404} (\bibinfo {year} {2010})}\BibitemShut {NoStop}%
\bibitem [{\citenamefont {Keldysh}(1965)}]{Keldysh1965}%
  \BibitemOpen
  \bibfield  {author} {\bibinfo {author} {\bibfnamefont {L.~V.}\ \bibnamefont
  {Keldysh}},\ }\bibfield  {title} {\bibinfo {title} {Diagram technique for non
  equilibrium processes},\ }\href@noop {} {\bibfield  {journal} {\bibinfo
  {journal} {Soviet Physics JETP}\ }\textbf {\bibinfo {volume} {20}},\ \bibinfo
  {pages} {1018} (\bibinfo {year} {1965})}\BibitemShut {NoStop}%
\bibitem [{\citenamefont {Wang}\ and\ \citenamefont
  {Sham}(2013)}]{wang2013quantum}%
  \BibitemOpen
  \bibfield  {author} {\bibinfo {author} {\bibfnamefont {Y.}~\bibnamefont
  {Wang}}\ and\ \bibinfo {author} {\bibfnamefont {L.~J.}\ \bibnamefont
  {Sham}},\ }\bibfield  {title} {\bibinfo {title} {Quantum approach of
  mesoscopic magnet dynamics with spin transfer torque},\ }\href@noop {}
  {\bibfield  {journal} {\bibinfo  {journal} {Physical Review B}\ }\textbf
  {\bibinfo {volume} {87}},\ \bibinfo {pages} {174433} (\bibinfo {year}
  {2013})}\BibitemShut {NoStop}%
\bibitem [{\citenamefont {Rammer}(2007)}]{rammer2007quantum}%
  \BibitemOpen
  \bibfield  {author} {\bibinfo {author} {\bibfnamefont {J.}~\bibnamefont
  {Rammer}},\ }\href@noop {} {\emph {\bibinfo {title} {Quantum field theory of
  non-equilibrium states}}}\ (\bibinfo  {publisher} {Cambridge University
  Press},\ \bibinfo {year} {2007})\BibitemShut {NoStop}%
\bibitem [{\citenamefont {Kamenev}\ and\ \citenamefont
  {Levchenko}(2009)}]{kamenev2009keldysh}%
  \BibitemOpen
  \bibfield  {author} {\bibinfo {author} {\bibfnamefont {A.}~\bibnamefont
  {Kamenev}}\ and\ \bibinfo {author} {\bibfnamefont {A.}~\bibnamefont
  {Levchenko}},\ }\bibfield  {title} {\bibinfo {title} {Keldysh technique and
  non-linear $\sigma$-model: basic principles and applications},\ }\href@noop
  {} {\bibfield  {journal} {\bibinfo  {journal} {Advances in Physics}\ }\textbf
  {\bibinfo {volume} {58}},\ \bibinfo {pages} {197} (\bibinfo {year}
  {2009})}\BibitemShut {NoStop}%
\bibitem [{\citenamefont {Kamenev}(2011)}]{kamenev2011field}%
  \BibitemOpen
  \bibfield  {author} {\bibinfo {author} {\bibfnamefont {A.}~\bibnamefont
  {Kamenev}},\ }\href@noop {} {\emph {\bibinfo {title} {Field theory of
  non-equilibrium systems}}}\ (\bibinfo  {publisher} {Cambridge University
  Press},\ \bibinfo {year} {2011})\BibitemShut {NoStop}%
\bibitem [{\citenamefont {Holstein}\ and\ \citenamefont
  {Primakoff}(1940)}]{holstein1940field}%
  \BibitemOpen
  \bibfield  {author} {\bibinfo {author} {\bibfnamefont {T.}~\bibnamefont
  {Holstein}}\ and\ \bibinfo {author} {\bibfnamefont {H.}~\bibnamefont
  {Primakoff}},\ }\bibfield  {title} {\bibinfo {title} {Field dependence of the
  intrinsic domain magnetization of a ferromagnet},\ }\href@noop {} {\bibfield
  {journal} {\bibinfo  {journal} {Physical Review}\ }\textbf {\bibinfo {volume}
  {58}},\ \bibinfo {pages} {1098} (\bibinfo {year} {1940})}\BibitemShut
  {NoStop}%
\bibitem [{Note1()}]{Note1}%
  \BibitemOpen
  \bibinfo {note} {The linearization is obtained by means of the
  Hubbard-Stratonovich transformation: \begin {equation*} e^{-\protect \frac
  {a}{2}\protect \tmspace +\thinmuskip {.1667em} x^2}= \protect \sqrt {\protect
  \frac {1}{2\protect \tmspace +\thinmuskip {.1667em} \pi \protect \tmspace
  +\thinmuskip {.1667em} a}} \DOTSI \intop \ilimits@ dI\protect \tmspace
  +\thinmuskip {.1667em} e^{-\protect \frac {I^2}{2\protect \tmspace
  +\thinmuskip {.1667em} a}-i\protect \tmspace +\thinmuskip {.1667em} x\protect
  \tmspace +\thinmuskip {.1667em} I} \end {equation*}}\BibitemShut {NoStop}%
\bibitem [{\citenamefont {Martin}\ \emph {et~al.}(1973)\citenamefont {Martin},
  \citenamefont {Siggia},\ and\ \citenamefont {Rose}}]{Martin1973}%
  \BibitemOpen
  \bibfield  {author} {\bibinfo {author} {\bibfnamefont {P.~C.}\ \bibnamefont
  {Martin}}, \bibinfo {author} {\bibfnamefont {E.~D.}\ \bibnamefont {Siggia}},\
  and\ \bibinfo {author} {\bibfnamefont {H.~A.}\ \bibnamefont {Rose}},\
  }\bibfield  {title} {\bibinfo {title} {Statistical dynamics of classical
  systems},\ }\href@noop {} {\bibfield  {journal} {\bibinfo  {journal}
  {Physical Review A}\ }\textbf {\bibinfo {volume} {8}},\ \bibinfo {pages}
  {423} (\bibinfo {year} {1973})}\BibitemShut {NoStop}%
\bibitem [{\citenamefont {Wang}\ and\ \citenamefont
  {Sham}(2012)}]{wang2012quantum}%
  \BibitemOpen
  \bibfield  {author} {\bibinfo {author} {\bibfnamefont {Y.}~\bibnamefont
  {Wang}}\ and\ \bibinfo {author} {\bibfnamefont {L.~J.}\ \bibnamefont
  {Sham}},\ }\bibfield  {title} {\bibinfo {title} {Quantum dynamics of a
  nanomagnet driven by spin-polarized current},\ }\href@noop {} {\bibfield
  {journal} {\bibinfo  {journal} {Physical Review B}\ }\textbf {\bibinfo
  {volume} {85}},\ \bibinfo {pages} {092403} (\bibinfo {year}
  {2012})}\BibitemShut {NoStop}%
\bibitem [{\citenamefont {Brown~Jr}(1963)}]{brown1963thermal}%
  \BibitemOpen
  \bibfield  {author} {\bibinfo {author} {\bibfnamefont {W.~F.}\ \bibnamefont
  {Brown~Jr}},\ }\bibfield  {title} {\bibinfo {title} {Thermal fluctuations of
  a single-domain particle},\ }\href@noop {} {\bibfield  {journal} {\bibinfo
  {journal} {Physical Review}\ }\textbf {\bibinfo {volume} {130}},\ \bibinfo
  {pages} {1677} (\bibinfo {year} {1963})}\BibitemShut {NoStop}%
\bibitem [{\citenamefont {Joos}\ and\ \citenamefont
  {Zeh}(1985)}]{joos1985emergence}%
  \BibitemOpen
  \bibfield  {author} {\bibinfo {author} {\bibfnamefont {E.}~\bibnamefont
  {Joos}}\ and\ \bibinfo {author} {\bibfnamefont {H.~D.}\ \bibnamefont {Zeh}},\
  }\bibfield  {title} {\bibinfo {title} {The emergence of classical properties
  through interaction with the environment},\ }\href@noop {} {\bibfield
  {journal} {\bibinfo  {journal} {Zeitschrift f{\"u}r Physik B Condensed
  Matter}\ }\textbf {\bibinfo {volume} {59}},\ \bibinfo {pages} {223} (\bibinfo
  {year} {1985})}\BibitemShut {NoStop}%
\bibitem [{\citenamefont {Zurek}(2003)}]{zurek2003decoherence}%
  \BibitemOpen
  \bibfield  {author} {\bibinfo {author} {\bibfnamefont {W.~H.}\ \bibnamefont
  {Zurek}},\ }\bibfield  {title} {\bibinfo {title} {Decoherence and the
  transition from quantum to classical--revisited},\ }\href@noop {} {\bibfield
  {journal} {\bibinfo  {journal} {arXiv preprint quant-ph/0306072}\ } (\bibinfo
  {year} {2003})}\BibitemShut {NoStop}%
\bibitem [{\citenamefont {Breuer}\ \emph {et~al.}(2002)\citenamefont {Breuer},
  \citenamefont {Petruccione} \emph {et~al.}}]{breuer2002theory}%
  \BibitemOpen
  \bibfield  {author} {\bibinfo {author} {\bibfnamefont {H.-P.}\ \bibnamefont
  {Breuer}}, \bibinfo {author} {\bibfnamefont {F.}~\bibnamefont {Petruccione}},
  \emph {et~al.},\ }\href@noop {} {\emph {\bibinfo {title} {The theory of open
  quantum systems}}}\ (\bibinfo  {publisher} {Oxford University Press on
  Demand},\ \bibinfo {year} {2002})\BibitemShut {NoStop}%
\bibitem [{\citenamefont {Nazarov}\ and\ \citenamefont
  {Blanter}(2009)}]{nazarov2009quantum}%
  \BibitemOpen
  \bibfield  {author} {\bibinfo {author} {\bibfnamefont {Y.~V.}\ \bibnamefont
  {Nazarov}}\ and\ \bibinfo {author} {\bibfnamefont {Y.~M.}\ \bibnamefont
  {Blanter}},\ }\href@noop {} {\emph {\bibinfo {title} {Quantum transport:
  introduction to nanoscience}}}\ (\bibinfo  {publisher} {Cambridge University
  Press},\ \bibinfo {year} {2009})\BibitemShut {NoStop}%
\bibitem [{Note2()}]{Note2}%
  \BibitemOpen
  \bibinfo {note} {For example, if we consider that electron are bounded in a
  region of $\bar {x}$ with dimension $L$ and with periodic boundary
  conditions, we have $N=1/\protect \sqrt {L}$; if $L$ is much greater with
  respect to the characteristic electron wave length, we may consider the
  continuous limit for $k$ and $N=1/\protect \sqrt {2\protect \tmspace
  +\thinmuskip {.1667em} \pi }$.}\BibitemShut {Stop}%
\bibitem [{Note3()}]{Note3}%
  \BibitemOpen
  \bibinfo {note} {This can take into account the action of a potential
  difference between the left and right regions: \begin {equation} \mu ^{s',d}
  = \epsilon _F + e\protect \tmspace +\thinmuskip {.1667em} V_0^d + s'\protect
  \tmspace +\thinmuskip {.1667em} 2\protect \tmspace +\thinmuskip {.1667em} \mu
  _B\protect \tmspace +\thinmuskip {.1667em} B^d_0, \end {equation} where
  $\epsilon _F$ is the Fermi energy, $V_0^d$ is an electric potential, $\mu _B$
  is the Bohr magneton, and $B^d_0$ is a local field due to the presence of
  hard ferromagnets layers (see Fig.~\ref {fig:modello}).}\BibitemShut {Stop}%
\bibitem [{Note4()}]{Note4}%
  \BibitemOpen
  \bibinfo {note} {In the continuous limit for $k$ (that is the linear
  dimension of the system along $\bar {x}$ is much greater with respect to the
  characteristic electron wavelength) and in the low temperature limit, it is
  possible to use the Sommerfeld expansion. In particular \begin {equation*}
  n_F^\mu (\epsilon )=\theta (\epsilon -\mu )\simeq \theta (\epsilon -\epsilon
  _F)+ \delta (\epsilon -\epsilon _F)\protect \tmspace +\thinmuskip {.1667em}
  (\mu -\epsilon _F), \end {equation*} in the zero temperature limit and
  assuming that all the chemical potentials have similar values: $\mu ^{s'
  d}\simeq \epsilon _F$}\BibitemShut {NoStop}%
\bibitem [{Note5()}]{Note5}%
  \BibitemOpen
  \bibinfo {note} {Observe that, in our expression~\protect \textup {\hbox
  {\mathsurround \z@ \protect \normalfont (\ignorespaces \ref
  {eqn:classico-quanto-azione}\unskip \@@italiccorr )}}, $D_0^{A/R}$ is an
  addend of $-i\protect \tmspace +\thinmuskip {.1667em}[G^{-1}]^{A/R}$. In
  particular, the fact that $\Re D^{A}_0=-\Re D^{R}_0$ and $\Im D^{A}_0=\Im
  D^{R}_0$ guarantees that the action component $S_{\protect \text {cl-q}}$ is
  real.}\BibitemShut {Stop}%
\bibitem [{Note6()}]{Note6}%
  \BibitemOpen
  \bibinfo {note} {A similar approximation is made in~\cite {wang2013quantum},
  since only one electron scattering per time is considered.}\BibitemShut
  {Stop}%
\bibitem [{Note7()}]{Note7}%
  \BibitemOpen
  \bibinfo {note} {In particular, for sufficiently small values of $\omega $,
  we have that the the contribution to $S(\omega )$ is non negligeable only for
  $\epsilon _{\delimiter 69640972 k_1\delimiter 86418188 } \sim \epsilon
  _{\delimiter 69640972 k_2\delimiter 86418188 }$; but in that case, for
  temperatures and differential potentials sufficiently small, $n_F^{s_1'
  d_1}(\epsilon _{\delimiter 69640972 k_1\delimiter 86418188 }) -n_F^{s_2'
  d_1}(\epsilon _{\delimiter 69640972 k_2\delimiter 86418188 } )$ is non zero
  only for $\epsilon _{\delimiter 69640972 k_1\delimiter 86418188 } \sim
  \epsilon _{\delimiter 69640972 k_2\delimiter 86418188 }\sim \epsilon _F$. So
  we have to assume $\epsilon _F\gg \omega $.}\BibitemShut {Stop}%
\bibitem [{\citenamefont {{\"O}ttinger}(1996)}]{ottinger1996stochastic}%
  \BibitemOpen
  \bibfield  {author} {\bibinfo {author} {\bibfnamefont {H.}~\bibnamefont
  {{\"O}ttinger}},\ }\href {https://books.google.it/books?id=WvJIAQAAIAAJ}
  {\emph {\bibinfo {title} {Stochastic Processes in Polymeric Fluids: Tools and
  Examples for Developing Simulation Algorithms}}}\ (\bibinfo  {publisher}
  {Springer},\ \bibinfo {year} {1996})\BibitemShut {NoStop}%
\bibitem [{\citenamefont {Chandrasekhar}(1943)}]{chandrasekhar1943stochastic}%
  \BibitemOpen
  \bibfield  {author} {\bibinfo {author} {\bibfnamefont {S.}~\bibnamefont
  {Chandrasekhar}},\ }\bibfield  {title} {\bibinfo {title} {Stochastic problems
  in physics and astronomy},\ }\href@noop {} {\bibfield  {journal} {\bibinfo
  {journal} {Reviews of Modern Physics}\ }\textbf {\bibinfo {volume} {15}},\
  \bibinfo {pages} {1} (\bibinfo {year} {1943})}\BibitemShut {NoStop}%
\end{thebibliography}%

\end{document}